%% file: main.tex
\documentclass[acmsmall]{acmart}

\acmJournal{PACMPL}
\acmVolume{1}
\acmNumber{CONF} 
\acmArticle{1}
\acmYear{2018}
\acmMonth{1}
\acmDOI{} 
\startPage{1}

\setcopyright{none}

\usepackage[utf8]{inputenc} 
\usepackage[T1]{fontenc}    
\usepackage{hyperref}       
\usepackage{url}            
\usepackage{booktabs}       
\usepackage{amsfonts}       
\usepackage{nicefrac}       
\usepackage{microtype}      
\usepackage{graphicx}
\usepackage{xspace}
\usepackage{amsmath}
\usepackage{amssymb}
\usepackage{multirow}
\usepackage{adjustbox}
\usepackage{listings}
\usepackage{subcaption} 
\usepackage{tikz}
\usepackage{xcolor}
\usepackage{flushend}
\usepackage{textcomp}

\theoremstyle{definition}
\newtheorem{definition}{Definition}[section]

\newcommand{\cf}{\hbox{\emph{cf.}}\xspace}
\newcommand{\etal}{\hbox{\emph{et al.}}\xspace}
\newcommand{\eg}{\hbox{\emph{e.g.}}\xspace}
\newcommand{\ie}{\hbox{\emph{i.e.}}\xspace}

\newcommand{\wrt}{\hbox{\emph{w.r.t.}}\xspace}
\newcommand{\etc}{\hbox{\emph{etc.}}\xspace}
\newcommand{\resp}{\hbox{\emph{resp.}}\xspace}

\newcommand{\dypro}{DYPRO\xspace}
\newcommand{\coset}{COSET\xspace}
\newcommand{\liger}{\textsc{LiGer}\xspace}

\DeclareMathOperator*{\argmax}{argmax}

\makeatletter
\newenvironment{btHighlight}[1][]
{\begingroup\tikzset{bt@Highlight@par/.style={#1}}\begin{lrbox}{\@tempboxa}}
	{\end{lrbox}\bt@HL@box[bt@Highlight@par]{\@tempboxa}\endgroup}

\newcommand\btHL[1][]{%
	\begin{btHighlight}[#1]\bgroup\aftergroup\bt@HL@endenv%
	}
	\def\bt@HL@endenv{%
	\end{btHighlight}%
	\egroup
}
\newcommand{\bt@HL@box}[2][]{%
	\tikz[#1]{%
		\pgfpathrectangle{\pgfpoint{1pt}{0pt}}{\pgfpoint{\wd #2}{\ht #2}}%
		\pgfusepath{use as bounding box}%
		\node[anchor=base west, fill=orange!25,outer sep=.5pt,inner xsep=0.5pt, inner ysep=-0.3pt, rounded corners=2pt, minimum height=\ht\strutbox-.1pt,#1]{\raisebox{.01pt}{\strut}\strut\usebox{#2}};
	}%
}
\makeatother

\definecolor{codegreen}{rgb}{0,0.6,0}
\definecolor{codegray}{rgb}{0.5,0.5,0.5}
\definecolor{codepurple}{rgb}{0.58,0,0.82}

\lstdefinestyle{mystyle}{
	frame=single,  
	commentstyle=\color{codegreen},
	keywordstyle=\color{blue}\bfseries,
	numberstyle=\tiny\color{codegray},
	stringstyle=\color{codepurple},
	basicstyle=\linespread{0.85}\fontsize{5.5}{8.8}\ttfamily\bfseries,
	breakatwhitespace=false,         
	breaklines=false,                 
	captionpos=b,                    
	keepspaces=true,                 
	numbers=none,                    
	numbersep=3pt,                  
	showspaces=false,                
	showstringspaces=false,
	showtabs=false,                  
	tabsize=2,
	language=[Sharp]C,
	moredelim=**[is][\btHL]{`}{`},
	moredelim=**[is][{\btHL[fill=red!40]}]{@}{@},
}

\begin{document}

\title{Learning Blended, Precise Semantic Program Embeddings}         


\author{Ke Wang}
\authornote{work done before joining Visa Research.}          
\orcid{nnnn-nnnn-nnnn-nnnn}             
\affiliation{
  \position{Position1}
  \institution{Visa Research}            
  \country{U.S.A}                    
}
\email{kewang@visa.com}          

\author{Zhendong Su}
\orcid{nnnn-nnnn-nnnn-nnnn}             
\affiliation{
  \department{Department of Computer Science}             
  \institution{ETH Zurich}           
  \country{Switzerland}                   
}
\email{zhendong.su@inf.ethz.ch}         

\input{abstract}

\ccsdesc[500]{Software and its engineering~General programming languages}
\ccsdesc[300]{Social and professional topics~History of programming languages}


\maketitle

\input{intro}

\input{over}

\input{pre}

\input{method}

\input{impl}
\input{eva}

\input{related}
\input{conc}


\bibliography{references}

%

\end{document}

%% file: abstract.tex
\begin{abstract}
	Learning neural program embeddings is key to utilizing deep
        neural networks in program languages research --- precise and
        efficient program representations enable the application of
        deep models to a wide range of program analysis tasks.
        Existing approaches predominately learn to embed programs from
        their source code, and, as a result, they do not capture deep,
        precise program semantics. On the other hand, models learned
        from runtime information critically depend on the quality of
        program executions, thus leading to trained models with highly
        variant quality. This paper tackles these inherent weaknesses
        of prior approaches by introducing a new deep neural network,
        \liger, which learns program representations from a mixture of
        symbolic and concrete execution traces. We have evaluated
        \liger on \coset, a recently proposed benchmark suite for
        evaluating neural program embeddings. Results show \liger (1)
        is significantly more accurate than the state-of-the-art
        syntax-based models Gated Graph Neural Network and code2vec in
        classifying program semantics, and (2) requires on average 10x
        fewer executions covering 74\% fewer paths than the
        state-of-the-art dynamic model \dypro. Furthermore, we extend
        \liger to predict the name for a method from its body's vector
        representation.  Learning on the same set of functions (more
        than 170K in total), \liger significantly outperforms
        code2seq, the previous state-of-the-art for method name
        prediction. 

%

\end{abstract}

%% file: intro.tex
\section{Introduction}
\label{sec:intro}

Learning representations has been a major focus in deep learning
research for the past several years. Mikolov~\etal pioneered the field
with their seminal work on learning word
embeddings~\cite{Mikolov:2013,mikolov2013efficient}.  The idea is to
construct a vector space for a corpus of text such that words found in 
similar contexts in the corpus are located in close proximity to one
another in the vector space. Word embeddings, along with other
representation learning (\eg doc2vec~\cite{le:2014}), become 
vital in solving many downstream Natural Language Processing (NLP)
tasks such as language modeling~\cite{Bengio:2003:NPL:944919.944966}
and sentiment classification~\cite{Glorot:2011:DAL:3104482.3104547}.

Similar to word embeddings, the goal of this paper is to learn
\emph{program embeddings}, vector representations of program
semantics. By learning program embeddings, the power of deep neural
networks (DNNs) can be utilized to tackle many program analysis
tasks. For example, \citet{Alon:2019:CLD:3302515.3290353} presents a
DNN to predict the name of a method given its
body. \citet{wang2017dynamic} propose a deep model to guide the
repair of student programs in Massive Open Online Courses
(MOOCs). Despite such notable advances, an important challenge
remains: How to tackle the precision and efficiency issues in learning
program embeddings? As illustrated by~\citet{wang2019coset}, due to
the inherent gap between program syntax and semantics, models learned
from source code (\ie, the \textit{static models}) can be 
imprecise at capturing semantic properties. Consider for example the
programs in Figure~\ref{fig:exam}. State-of-the-art static models can
neither recognize the equivalent semantics between programs in
Figures~\ref{fig:a} and~\ref{fig:c}, nor the different semantics
between programs in Figures~\ref{fig:a} and~\ref{fig:b}. The reason
for this is quite obvious --- static models base their predictions on
the surface-level program syntax. Specifically, programs~\ref{fig:a}
and~\ref{fig:b} are syntactically much more similar than
programs~\ref{fig:a} and~\ref{fig:c} despite that programs~\ref{fig:a}
and~\ref{fig:c} implement the same sorting
strategy, namely \textit{Bubble Sort}.

\begin{figure*}[htbp!]
	\vskip 0pt
	\begin{subfigure}{0.32\textwidth}
		\lstset{style=mystyle}
\begin{lstlisting}[linewidth=4.24cm]
static int SortI(int[] A)
{
    int left = 0;
    int right = @A.Length - 1@;

    for(int i=@right@;i@>left@;i@--@) { 
        for(int j=@left@;j@<i@;j@++@) { 
           if (A[j] > A[j + 1]) { 
                int tmp = A[j];
                A[j] = A[j + 1];
                A[j + 1] = tmp;
    }}}

    return A;
}
\end{lstlisting}
		\caption{Bubble Sort}
		\label{fig:a}
	\end{subfigure}
	\begin{subfigure}{0.32\textwidth}	
		\lstset{style=mystyle}
\begin{lstlisting}[linewidth=4.25cm]
static int SortII(int[] A)
{
   int left = 0;
   int right = @A.Length@;

   for(int i=@left@;i@<right@;i@++@) {
      for(int j=@i-1@;j@>=left@;j@--@) { 
          if (A[j] > A[j + 1]) { 
             int tmp = A[j];
             A[j] = A[j + 1];
             A[j + 1] = tmp;
   }}}

   return A;
}

\end{lstlisting}
		\caption{Insertion Sort}
		\label{fig:b}
	\end{subfigure}
	\begin{subfigure}{0.32\textwidth}
		\lstset{style=mystyle}
\begin{lstlisting}
static int SortIII(int[] A)
{
   int swappbit = 1;
   while (swappbit != 0) { 
      swappbit = 0;
      for (int i=0;i<A.Length-1;i++)
      { 
          if (A[i + 1] > A[i]) { 
              int tmp = A[i];
              A[i] = A[i + 1];
              A[i + 1] = tmp;
              swappbit = 1;
      }}}
   return A;
}
\end{lstlisting}		
		\caption{Bubble Sort}
		\label{fig:c}
	\end{subfigure}
	\caption{Example programs that implement a sorting
          routine. Code highlighted within the shadow boxes depicts
          the syntactic differences between programs~\ref{fig:a}
          and~\ref{fig:b}.}
	\label{fig:exam}
\end{figure*}

\begin{figure*}[htbp!]
	\vskip 0pt
	\begin{subfigure}{0.35\textwidth}
		\lstset{style=mystyle}
\begin{lstlisting}[basicstyle=\linespread{0.85}\fontsize{7.3}{8.8}\ttfamily,linewidth=4.66cm,escapechar=^]
^\text{$\{$A:[8, 5, 1, 4, 3]; \textbf{left}:\textbf{0}; right:$\bot$$\}$}^
^\text{$\{$A:[8, 5, 1, 4, 3]; left:0; \textbf{right}:\textbf{4}$\}$}^
^\text{$\{$\textbf{A}:[\textbf{5}, 5, 1, 4, 3]; left:0; right:4$\}$}^
^\text{$\{$\textbf{A}:[5, \textbf{8}, 1, 4, 3]; left:0; right:4$\}$}^
^\text{$\{$\textbf{A}:[5, \textbf{1}, 1, 4, 3]; left:0; right:4$\}$}^
^\text{$\{$\textbf{A}:[5, 1, \textbf{8}, 4, 3]; left:0; right:4$\}$}^
^\text{$\{$\underline{\textbf{A}:[5, 1, \textbf{4}, 4, 3]}; left:0; right:4$\}$}^
^\text{$\{$\underline{\textbf{A}:[5, 1, 4, \textbf{8}, 3]}; left:0; right:4$\}$}^
^\text{$\{$\underline{\textbf{A}:[5, 1, 4, \textbf{3}, 3]}; left:0; right:4$\}$}^
^\text{$\{$\underline{\textbf{A}:[5, 1, 4, 3, \textbf{8}]}; left:0; right:4$\}$}^
^\text{$\{$\underline{\textbf{A}:[\textbf{1}, 1, 4, 3, 8]}; left:0; right:4$\}$}^
^\text{$\{$\underline{\textbf{A}:[1, \textbf{5}, 4, 3, 8]}; left:0; right:4$\}$}^
^\text{$\{$\underline{\textbf{A}:[1, \textbf{4}, 4, 3, 8]}; left:0; right:4$\}$}^
^\text{$\{$\underline{\textbf{A}:[1, 4, \textbf{5}, 3, 8]}; left:0; right:4$\}$}^
^\text{$\{$\textbf{A}:[1, 4, \textbf{3}, 3, 8]; left:0; right:4$\}$}^
^\text{$\{$\textbf{A}:[1, 4, 3, \textbf{5}, 8]; left:0; right:4$\}$}^
^\text{$\{$\textbf{A}:[1, \textbf{3}, 3, 5, 8]; left:0; right:4$\}$}^
^\text{$\{$\textbf{A}:[1, 3, \textbf{4}, 5, 8]; left:0; right:4$\}$}^
\end{lstlisting}
		\caption{}
		\label{fig:exea}
	\end{subfigure}
	\begin{subfigure}{0.35\textwidth}	
		\lstset{style=mystyle}
\begin{lstlisting}[basicstyle=\linespread{0.85}\fontsize{7.3}{8.8}\ttfamily,linewidth=4.66cm,escapechar=^]
^\text{$\{$A:[8, 5, 1, 4, 3]; \textbf{left}:\textbf{0}; right:$\bot$$\}$}^
^\text{$\{$A:[8, 5, 1, 4, 3]; left:0; \textbf{right}:\textbf{5}$\}$}^
^\text{$\{$\textbf{A}:[\textbf{5}, 5, 1, 4, 3]; left:0; right:5$\}$}^
^\text{$\{$\textbf{A}:[5, \textbf{8}, 1, 4, 3]; left:0; right:5$\}$}^
^\text{$\{$\textbf{A}:[5, \textbf{1}, 1, 4, 3]; left:0; right:5$\}$}^
^\text{$\{$\textbf{A}:[5, 1, \textbf{8}, 4, 3]; left:0; right:5$\}$}^
^\text{$\{$\underline{\textbf{A}:[\textbf{1}, 1, 8, 4, 3]}; left:0; right:5$\}$}^
^\text{$\{$\underline{\textbf{A}:[1, \textbf{5}, 8, 4, 3]}; left:0; right:5$\}$}^
^\text{$\{$\underline{\textbf{A}:[1, 5, \textbf{4}, 4, 3]}; left:0; right:5$\}$}^
^\text{$\{$\underline{\textbf{A}:[1, 5, 4, \textbf{8}, 3]}; left:0; right:5$\}$}^
^\text{$\{$\underline{\textbf{A}:[1, \textbf{4}, 4, 8, 3]}; left:0; right:5$\}$}^
^\text{$\{$\underline{\textbf{A}:[1, 4, \textbf{5}, 8, 3]}; left:0; right:5$\}$}^
^\text{$\{$\underline{\textbf{A}:[1, 4, 5, \textbf{3}, 3]}; left:0; right:5$\}$}^
^\text{$\{$\underline{\textbf{A}:[1, 4, 5, 3, \textbf{8}]}; left:0; right:5$\}$}^
^\text{$\{$\textbf{A}:[1, 4, \textbf{3}, 3, 8]; left:0; right:5$\}$}^
^\text{$\{$\textbf{A}:[1, 4, 3, \textbf{5}, 8]; left:0; right:5$\}$}^
^\text{$\{$\textbf{A}:[1, \textbf{3}, 3, 5, 8]; left:0; right:5$\}$}^
^\text{$\{$\textbf{A}:[1, 3, \textbf{4}, 5, 8]; left:0; right:5$\}$}^
\end{lstlisting}
		\caption{}
		\label{fig:exeb}
	\end{subfigure}
	\begin{subfigure}{0.25\textwidth}
		\lstset{style=mystyle}
\begin{lstlisting}[basicstyle=\linespread{0.902}\fontsize{7.3}{8.8}\ttfamily,linewidth=3.8cm,escapechar=^]
^\text{$\{$A:[8, 5, 1, 4, 3]; \textbf{swapbit}:\textbf{1}$\}$}^
^\text{$\{$A:[8, 5, 1, 4, 3]; \textbf{swapbit}:\textbf{0}$\}$}^
^\text{$\{$\textbf{A}:[\textbf{5}, 5, 1, 4, 3]; swapbit:1$\}$}^
^\text{$\{$\textbf{A}:[5, \textbf{8}, 1, 4, 3]; swapbit:1$\}$}^
^\text{$\{$\textbf{A}:[5, \textbf{1}, 1, 4, 3]; swapbit:1$\}$}^
^\text{$\{$\textbf{A}:[5, 1, \textbf{8}, 4, 3]; swapbit:1$\}$}^
^\text{$\{$\textbf{A}:[\textbf{1}, 1, 8, 4, 3]; swapbit:1$\}$}^
^\text{$\{$\textbf{A}:[1, \textbf{5}, 8, 4, 3]; swapbit:1$\}$}^
^\text{$\{$\textbf{A}:[1, 5, \textbf{4}, 4, 3]; swapbit:1$\}$}^
^\text{$\{$\textbf{A}:[1, 5, 4, \textbf{8}, 3]; swapbit:1$\}$}^
^\text{$\{$\textbf{A}:[1, \textbf{4}, 4, 8, 3]; swapbit:1$\}$}^
^\text{$\{$\textbf{A}:[1, 4, \textbf{5}, 8, 3]; swapbit:1$\}$}^
^\text{$\{$\textbf{A}:[1, 4, 5, \textbf{3}, 3]; swapbit:1$\}$}^
^\text{$\{$\textbf{A}:[1, 4, 5, 3, \textbf{8}]; swapbit:1$\}$}^
^\text{$\{$\textbf{A}:[1, 4, \textbf{3}, 3, 8]; swapbit:1$\}$}^
^\text{$\{$\textbf{A}:[1, 4, 3, \textbf{5}, 8]; swapbit:1$\}$}^
^\text{$\{$\textbf{A}:[1, \textbf{3}, 3, 5, 8]; swapbit:1$\}$}^
^\text{$\{$\textbf{A}:[1, 3, \textbf{4}, 5, 8]; swapbit:1$\}$}^
\end{lstlisting}
		\caption{}
		\label{fig:exec}
	\end{subfigure}
	\caption{Encoding the executions of the programs in
          Figure~\ref{fig:exam} with the input array \texttt{A =
            [8,5,1,4,3]}. At each step, the variable in bold is
          updated. Steps that are underlined illustrate the semantic
          differences between bubble sort (program~\ref{fig:a}) and
          insertion sort (program~\ref{fig:b}) concerning only the
          manipulation of the input array \texttt{A}. Note that we
          have omitted all loop induction variables, variable
          \texttt{tmp} in both programs~\ref{fig:a} and~\ref{fig:b} as
          well as some steps that update the variable \texttt{swapbit} in
          program~\ref{fig:c} to simply our presentation.  }
	\label{fig:exe}
\end{figure*}

In parallel, a separate class of models have been proposed that embed
programs from their concrete execution traces (\ie, the
\textit{dynamic models}). Compared to source code, program executions
capture accurate, deep program semantics, thus offering benefits
beyond static models that reason over syntactic
representations. Figure~\ref{fig:exe} shows the executions of the
three programs with an input array \texttt{A = [8,5,1,4,3]} according
to the state-based encoding proposed by~\citet{wang2017dynamic}.
Naturally, the semantic relationship among the three programs becomes
much clearer.  Despite their advantages, the performance of dynamic
models heavily depends on the quality of program executions. In
particular, dynamic models, similar to dynamic program analysis, can
suffer from insufficient code coverage. Even for each covered path,
dynamic models may need a large number of execution instances to
generalize, resulting in a lengthy and expensive training process.

To tackle those aforementioned issues of both strands of prior work,
we introduce a novel, \emph{blended} approach for learning precise and
efficient representations of program semantics.  Our insight is to
blend the respective strengths of static and dynamic models to
mitigate their respective weaknesses. To this end, we propose a
carefully designed blended model to learn a deep, precise semantic
representation. Different from dynamic models that consider only
program states created along an execution path, our blended model
incorporates the additional symbolic representation of each statement
(\ie \emph{symbolic trace}) whose execution leads to a corresponding
program state.

The benefits of blending these feature dimensions are twofold. First,
learning from symbolic program encodings is shown difficult for DNNs~\cite{wang2019coset,wang2019learning}.
Concrete program states, which give live illustrations of program
behavior, provide explanations to DNNs about each symbolic statement's
semantics.  As a result, models trained on the combined features
capture deeper semantic properties than symbolic traces alone (\cf
Section~\ref{subsubsec:static}). Second, a symbolic trace typically
generalizes a large number of concrete executions.  Therefore,
symbolic traces present high-level, general descriptions of program
meaning to DNNs. It is for this very reason that symbolic traces stand
out as the major feature dimension from which models generalize.  In
the presence of the symbolic feature dimension, DNNs deemphasize the
role of dynamic program features, leading to reduced demand on the
number of concrete executions. As another benefit of our blended
model, we observe that DNNs trained on both feature dimensions are
also more resilient to the varying diversity on program
executions. Indeed, when the path coverage on the targeted program is
systematically reduced, our blended network largely maintains its
accuracy, and thus has improved data reliance
(\cf Section~\ref{subsubsec:reli}).

We have realized our approach in a new DNN, \liger, and extensively
evaluated it. Using \coset~\cite{wang2019coset}, a recently
proposed benchmark suite for evaluating neural program embeddings, 
we find \liger achieves significantly better accuracy and
stability\footnote{The ability to preserve its prediction against
  natural code changes such as code optimization and software
  refactoring.} than Gated Graph Neural Network
(GGNN)~\cite{allamanis2017learning} and
code2vec~\cite{Alon:2019:CLD:3302515.3290353}, two of the most widely
applied static models in their respective problem domains. Compared to
\dypro~\cite{wang2019learning}, the state-of-the-art dynamic model in
learning program embeddings, \liger is more accurate even when
generalizing from on average almost 10x fewer executions that cover
74\% fewer code paths per program.

We also extend \liger to solve the method name prediction problem
studied by~\citet{Alon:2019:CLD:3302515.3290353}. Our dataset contains
174,922 functions extracted from logs of coding interviews conducted
by an IT company for testing a candidate's programming skills. Each
function is written by a candidate to solve an algorithmic question
(\eg, merging two sorted arrays, determining if a string is a
palindrome, detecting the presence of a cycle in a linked list, \etc).
All method names were provided by the interviewers for describing the
functionality of each method.
We strip away the method names for models to predict. Results show
that \liger significantly outperforms all competing deep neural
architectures, including code2seq~\cite{alon2018code2seq}, the
previous state-of-the-art.

We make the following main contributions:
\begin{itemize}
	\item We propose a novel, blended approach that combines
          static and dynamic program features for learning precise and
          efficient representations of program semantics.
	
	\item We realize our approach in \liger, which achieves
          significantly better accuracy and stability than GGNN and
          code2vec on \coset and requires far fewer program
          executions for both training and testing than \dypro.
	
	\item We extend \liger to solve the method name prediction
          problem. Results show that \liger also significantly outperforms
          code2seq, the previous state-of-the-art DNN.
	
	\item We present the details of our extensive evaluation of
          \liger, including an ablation study that analyzes the
          contributions of \liger's several crucial components to its
          overall performance.
\end{itemize}

%% file: over.tex
\section{Overview}

This section overviews our blended approach of learning program
embeddings. In particular, it explains how we address the deficiencies
of static and dynamic models. We begin by formalizing the notion of
execution traces and several pertinent concepts.

\begin{figure*}[htbp!]
	\begin{subfigure}{0.3\textwidth}
		\lstset{style=mystyle}
\begin{lstlisting}[linewidth=5.2cm]
public bool isRotateString(string A,string B)
{
   if (A.Length != B.Length)
        return false;

   for (int i = 1; i < A.Length; i++) 
   {         
       string tail=A.Substring(i,A.Length-i);
       string wrap=A.Substring(0, i);

       if (tail + wrap == B)
           return true;
   }

   return false;
}
\end{lstlisting}
		\caption{}
		\label{fig:exel}
	\end{subfigure}
	\begin{subfigure}{0.6\textwidth}	
		\begin{center}
			\centerline{\includegraphics[width=.9\columnwidth]{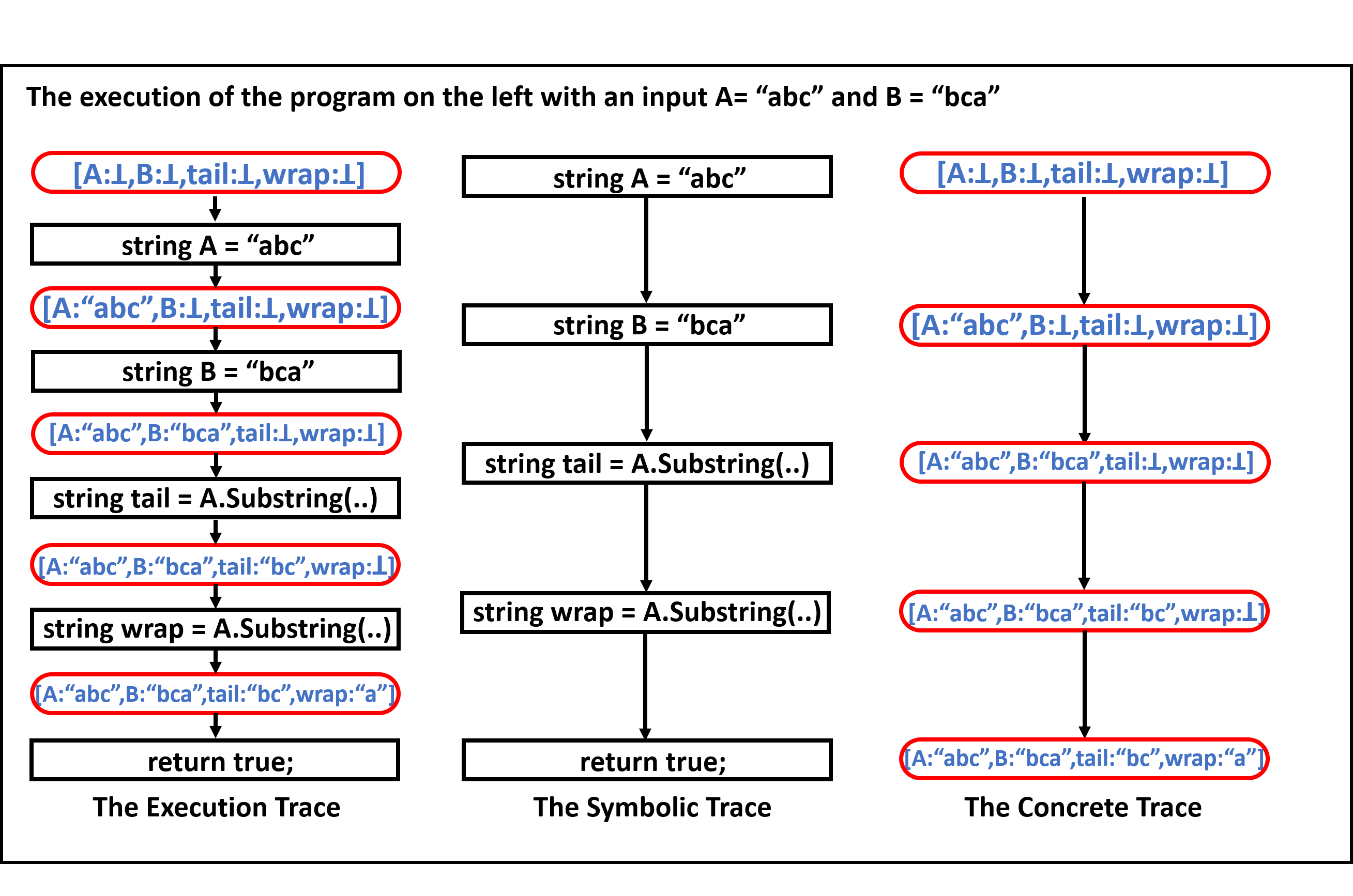}}
			\caption{}
		\label{fig:exer}
		\end{center}		
	\end{subfigure}
	\caption{Given the program in (\subref{fig:exel}), we give
          example execution traces, symbolic traces and concrete
          traces in (\subref{fig:exer}).  }
	\label{fig:traces}
\end{figure*}

\subsection{Formalization}

In general, given a program $P$ and an input $I$, an execution trace
is obtained by executing $P$ on $I$.  Its concept and notations are
standard, which we formalize more precisely below.

\begin{definition}(Execution Trace)
	An execution trace, denoted by $\pi$, is a sequence in the
        form of $s_{0} \rightarrow (e_{i}\rightarrow s_{i})^{*}$,
        where $e_{i}$ denotes a statement encountered as $P$ executes
        on an input $I$; $s_{i}$ denotes a program state, which is a
        set of variable/memory and value pairs immediately after the
        execution of statement $e_{i}$;  $s_{0}$ is the initial
        program state; and $^{*}$ denotes the Kleene star. 
\end{definition}

As an example, Figure~\ref{fig:exer} presents a graphical illustration
of an execution trace of the program in Figure~\ref{fig:exel} with
input \texttt{A = "abc"} and \texttt{B = "bca"}.

\begin{definition}(Symbolic Trace)
	Given an execution trace, $\pi$, a symbolic trace, 
	$\sigma$, is the sequence of statements visited 
	in $\pi$ in the form of $(e_{i}\rightarrow e_{i+1})^{*}$.	
\end{definition}

Similarly, Figure~\ref{fig:exer} also gives an example symbolic
trace, which is a projection of the execution trace \wrt the program
statements.

\begin{definition}(State Trace)
	Given an execution trace, $\pi$, a state trace, $\epsilon$, is
        the sequence of program states created in $\pi$ in the form of
        $(s_{i}\rightarrow s_{i+1})^{*}$.
\end{definition}

Again, Figure~\ref{fig:exer} shows an example state trace, which is a
projection of the execution trace \wrt the program states. 

\subsection{Motivation and Insight}

Learning program embeddings from execution traces has been explored in
the literature. The prior work can be divided into two categories:
static and dynamic. The former refers to learning program embeddings
exclusively on symbolic traces.  As an
example,~\citet{Henkel:2018:CVU:3236024.3236085} train a model from
symbolic traces for a code analogy task.
Although learning from symbolic traces captures, to a certain degree,
program properties more at the semantic level than purely the
syntactic source code, it suffers from the same fundamental issue of
all static models. That is such approaches leave the burden on the
deep models to reason about program semantics through a syntactic
representation, a task that is proven to be challenging even for
state-of-the-art DNNs. To give a simple example, a capable neural
network needs to recognize the identical semantics \texttt{i+=i} and
\texttt{i*=2} denote because such variations are ubiquitous in
real-world code. Note that Henkel~\etal's approach relies on
user-defined abstraction templates, which unfortunately do not address
the problem's root cause.

In contrast to the symbolic trace-based approaches, another line of
work only considers concrete state traces for learning program
representations. In particular,~\citet{wang2017dynamic} propose a
model learned from concrete state traces to predict the type of errors
students make in their programming assignments. The intuition behind
the approach is to capture the semantics of a program through the
states that are created in an execution. The advantage of their
approach is the canonicalization of syntactic variations as programs
of equivalent semantics will always create identical program states
regardless of their syntactic differences (\eg, the earlier example
involving \texttt{i+=i} and \texttt{i*=2}). Despite this strength,
models that embed programs from concrete state traces have their own
weaknesses. In principle, a symbolic trace can be instantiated to a
large or arbitrary number of concrete states traces.  Therefore,
symbolic traces lay the foundation of feature representations. By
completely disregarding the program syntax, deep models lose the
high-level overview of the execution trace, therefore demanding a
large number of concrete traces to compensate. Assuming that a model
requires $M$ concrete traces to estimate the semantics of one symbolic
trace, learning a program yielding $N$ total symbolic traces amounts
to $M*N$ concrete traces. This drastic increase in the amount of
training data leads to lengthy and inefficient training.

The deficiencies of the static and dynamic models together motivate
the design of \liger. By simply exposing the entire execution traces
(\ie both symbolic and concrete state traces) as structured inputs,
\liger combines the strengths of both types of models and outperforms ones
learned from either symbolic or concrete traces alone. On one hand, concrete
traces help \liger deal with the challenge of learning from symbolic
program representations. Instead of generalizing from high-level
program symbols, \liger is also provided with low-level concrete
explanations.  Consider the earlier example with \texttt{i+=i} and
\texttt{i*=2}.  Although the two statements are represented
differently in terms of symbolic traces, their identical program
states force \liger to inject the notion of equivalent semantics
between the two statements. Ultimately, this allows \liger's to reduce
the difficulty of reasoning about program semantics from syntax.

Additionally, since the symbolic representation of an execution
trace is still present in the feature representation, \liger has a 
general, symbolic view of the execution trace, therefore does not need
a large number of concrete traces to generalize. Thus, it needs
less training data. 

Through our extensive experiments, we observe that \liger possesses an
interesting benefit. As we systematically lower the path coverage of
programs in both the training and test sets, \liger is able to
maintain its accuracy. This property of \liger helps itself address
the intrinsic limitations of the dynamic models. That is, even when
programs are hard to cover, \liger can still reason about the
semantics of a program from the limited available traces, thus
further reducing its reliance on training data.

%% file: pre.tex
\section{Preliminaries}

This section reviews necessary background, in particular, recurrent
neural networks and attention neural networks, the building blocks of
\liger.

\subsection{Recurrent Neural Network}
\label{subsec:RNN}

A recurrent neural network (RNN) is a class of artificial neural
networks that are distinguished from feedforward networks by their
feedback loops.  This allows RNNs to ingest their own outputs as
inputs. It is often said that RNNs have memory, enabling them to
process sequences of inputs.

Here we briefly describe the computation model of a vanilla RNN. Given
an input sequence, embedded into a sequence of vectors $x = (x_{1},
\cdot\cdot\cdot, x_{T_{x}})$, an RNN with $N$ inputs, a single hidden
layer with $M$ hidden units, and $Q$ output units. We define the RNN's
computation as follows:
\begin{align}
h_{t} &= f(W * x_{t} + V * h_{t-1}) \label{equ:1}\\
o_{t} &= \mathit{softmax}(Z * h_{t}) \notag
\end{align}
where $x_{t} \in \mathbb{R}^{N}$, $h_{t} \in \mathbb{R}^{M}$,
$o_{t}\in \mathbb{R}^{Q}$ is the RNN's input, hidden state and output
at time $t$, $f$ is a non-linear function (\eg tanh or sigmoid), $W
\in \mathbb{R}^{M*N}$ denotes the weight matrix for connections from
input layer to hidden layer, $V \in \mathbb{R}^{M*M}$ is the weight
matrix for the recursive connections (\ie from hidden state to itself)
and $Z \in \mathbb{R}^{Q*M}$ is the weight matrix from hidden to the
output layer.

\subsection{Neural Attention Network}
\label{subsec:atte}

Before we describe attention neural networks, we give a brief overview
of the underlying framework --- \textit{Encoder-Decoder} --- proposed
by~\citet{cho2014learning} and~\citet{devlin2014fast}.

The encoder-decoder~\cite{cho2014learning,devlin2014fast} neural
architecture was first introduced in the field of machine translation.  An
encoder neural network reads and encodes a source sentence into a
vector based on which a decoder outputs a translation. From a
probabilistic point of view, the goal of translation is to find the
target sentence $L_{t}$ that maximizes the conditional probability of
$L_{t}$ given source sentence $L_{s}$ (\ie $\argmax_{L_{t}}\!
P(L_{t}|L_{s})$).

Using the terminologies defined in Section~\ref{subsec:RNN}, we
explain the computation model of an encoder-decoder. Given an input
sequence $x$, the encoder performs the computation defined in
Equation~\ref{equ:1} and spits out its final hidden state $c$ (\ie
$c=h_{T_{x}}$). The decoder is responsible for predicting each word
$y_{t}$ given the vector $c$ and all the previously predicted words
$(y_{1}, \cdot\cdot\cdot, y_{t-1})$. In other words, the decoder
outputs the probability distribution of $y = (y_{1}, \cdot\cdot\cdot,
y_{T_{y}})$ by decomposing the joint probability into the ordered
conditionals:
\begin{equation}
  P(y) = \prod_{t=1}^{T_{y}} P(y_{t} | (y_{1}, \cdot\cdot\cdot, y_{t-1}), c) \notag
\end{equation}
With an RNN, each ordered conditional is defined as: 
\begin{equation}
  P(y_{t} | (y_{1}, \cdot\cdot\cdot, y_{t-1}), c) = g(y_{t-1},d_{t},c) \notag \label{equ:4}
\end{equation}
where $d_{t}$ is the hidden state of the decoder RNN at time $t$.
 
An issue of this encoder–decoder architecture is that the encoder has
to compress all the information from a source sentence into a vector
to feed the decoder. This is especially problematic when the encoder has to
deal with long sentences. To address this issue,
\citet{bahdanau2014neural} introduced an attention mechanism on top of
the standard encoder-decoder framework that learns to align and
translate simultaneously. The proposed solution is to enable the
decoder network to search the most relevant
information from the source sentence to concentrate when decoding each
target word. In particular, instead of fixing each conditional
probability on the vector $c$ in Equation~\ref{equ:4}, a distinct
context vector $c_{t}$ for each $y_{t}$ is used:
\begin{equation}
  P(y_{t} | (y_{1}, \cdot\cdot\cdot, y_{t-1}), x) = g(y_{t-1},d_{t},c_{t}) \notag \label{equ:5}
\end{equation}

To compute the context vector $c_{t}$, a bi-directional RNN is adopted
which reads the input sequence $x$ from both directions (\ie, from
$x_{1}$ to $x_{T_{x}}$ and vice versa), and produces a sequence of
forward hidden states
($\overrightarrow{h_{1}},\cdot\cdot\cdot,\overrightarrow{h_{T_{x}}}$)
and backward hidden states
($\overleftarrow{h_{1}},\cdot\cdot\cdot,\overleftarrow{h_{T_{x}}}$).
We obtain an annotation $h_{d}$ for each word $x_{d}$ by concatenating
the forward hidden state $\overrightarrow{h_{d}}$ and the backward one
$\overleftarrow{h_{d}}$. Now $c_{t}$ can be computed as a weighted sum
of these annotations $h_{d}$ :
\begin{equation}
c_{t} = \sum_{d=1}^{T_{x}} \alpha_{\mathit{td}}h_{d}      \label{equ:6}
\end{equation}
The attention weight $\alpha_{\mathit{td}}$ of each annotation $h_{d}$ is computed by
\begin{equation}
\alpha_{\mathit{td}} = \frac{\mathit{exp}(\mu_{\mathit{td}})}{\sum_{k=1}^{T_{x}}\mathit{exp}(\mu_{\mathit{tk}})}   \notag \label{equ:7}
\end{equation}
where $\mu_{\mathit{td}} = a(d_{t-1},h_{d})$ 
is the attention score which reflects the importance of the annotation
$h_{d}$ \wrt the previous hidden state $d_{t-1}$ in deciding the next
state $d_{t}$ and generating $y_{t}$. The parameter $a$ stands for a
feedforward neural network that is jointly trained with the system's other
components.

%% file: method.tex
\section{Model}

This section presents the technical details of \liger's architecture
and discusses how to extend \liger to build an attention network to
solve the problem of predicting method names.

\begin{figure*}[t]
	\begin{center}
		\centerline{\includegraphics[width=.9\columnwidth]{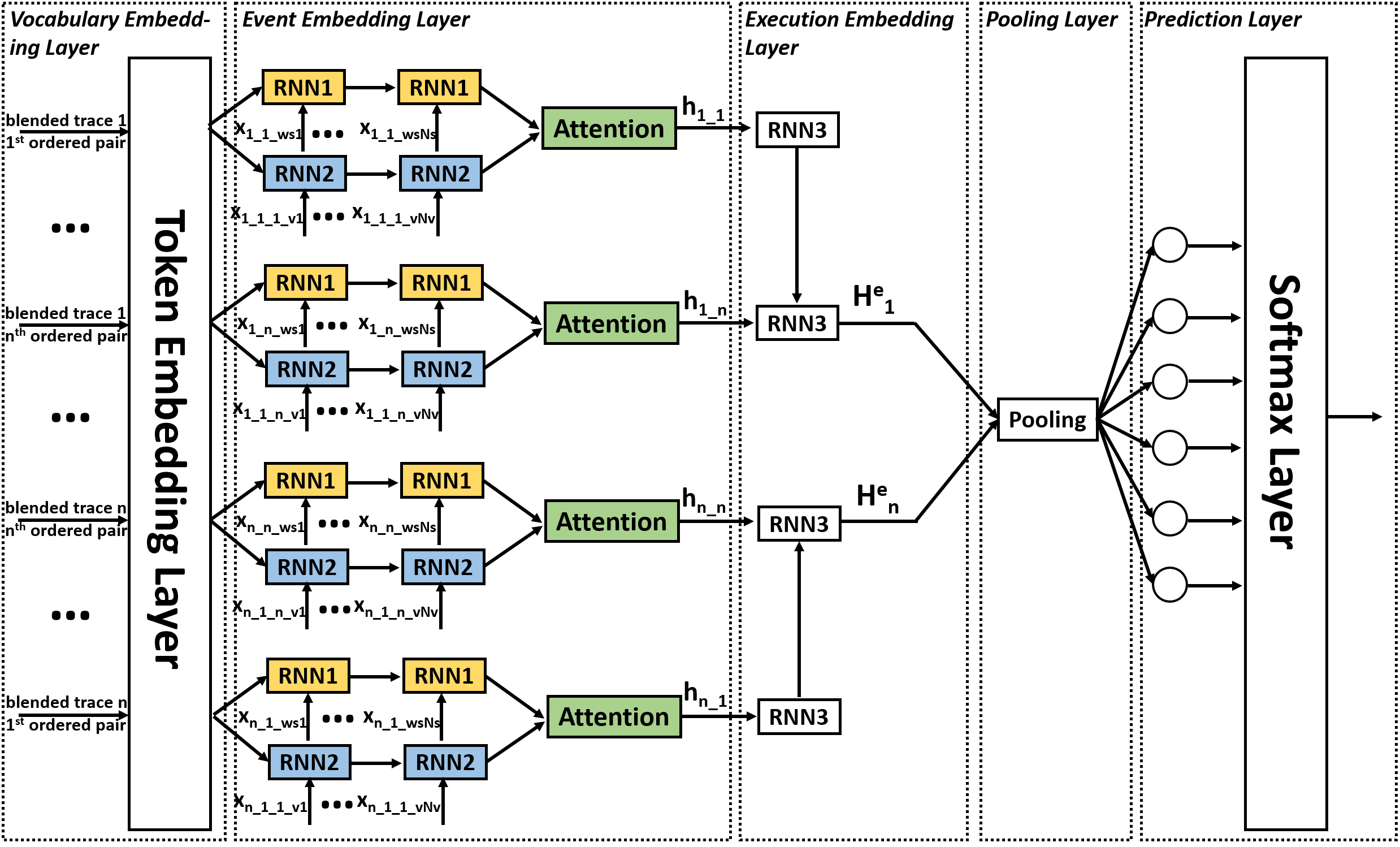}}
		\caption{\liger's architecture.}
		\label{fig:archi}
	\end{center}
\end{figure*}

\subsection{\liger's Architecture}

\liger's architecture is depicted in Figure~\ref{fig:archi}.  At a
high level, \liger uses three RNNs to encode an entire execution into
a vector. To elevate the learning at the execution level to the
program level, a pooling layer is designed to compress execution
embeddings into program embeddings. Below, we split the entire network
into five layers and discuss each layer in detail.

\vspace*{3pt}
\noindent
{\textbf{Terminology.}}
Given a program $P$, we denote its set of tokens as
$w_{P_{1}},\cdot\cdot\cdot,w_{P_{N_{p}}}$. Each token 
$w_{P_{i}} \in \mathcal{D}_{s}$ where $\mathcal{D}_{s}$ 
is the set of all tokens extracted from all programs in 
our dateset. Within $P$'s token set,
$w_{v_{1}},\cdot\cdot\cdot,w_{v_{N_{v}}}$ denote the variables.  To turn
$P$ in its source code form to the format \liger requires, we
symbolically execute $P$ to obtain $U$ distinct paths,
where each path $\sigma_{i}$ is associated with a condition
$\phi_{i}$.  By solving $\phi_{i}$, we obtain concrete traces 
$\epsilon_{i\_1},\cdot\cdot\cdot,\epsilon_{i\_N_{\epsilon}}$. We encode each
statement in $\sigma_{i}$ as a sequence of tokens
$[w_{s_{1}},\cdot\cdot\cdot,w_{s_{N_{s}}}]$, and each program state in 
a concrete trace, $\epsilon_{i\_i^\prime}$, as a tuple of values
$(v_{1},\cdot\cdot\cdot,v_{N_{v}})$ where $v_{i} \in \mathcal{D}_{d}$ is
the value of $w_{v_{i}}$. $\mathcal{D}_{d}$ refers to the set of all values 
any variable has ever been assigned in any concrete trace of any program 
in our dataset.
%

\vspace*{3pt}
\noindent
{\textbf{Vocabulary Embedding Layer.}}  In this layer, each $w_{i} \in
\mathcal{D}_{s}$ and $v_{i} \in \mathcal{D}_{d}$ will be assigned a
vector.  Consider the $j$-th statement in $\sigma_{i}$, after the
vocabulary embedding phase, the statement will be encoded as
$[x_{i\_j\_w_{s_{1}}},\cdot\cdot\cdot,x_{i\_j\_w_{s_{N_{s}}}}]$. 
Similarly, the state created by $\sigma_{i}$'s $j$-th statement in
$\epsilon_{i\_i^\prime}$ will be encoded as
$(x_{i\_i^\prime\_j\_v_{1}},\cdot\cdot\cdot,x_{i\_i^\prime\_j\_v_{N_{v}}})$.



\vspace*{3pt}
\noindent
{\textbf{Fusion Layer.}}  This layer embeds each statement in
$\sigma_{i}$ and each state in $\epsilon_{i\_i^\prime}$ before fusing
the two feature dimensions, which forms the core of our blended
approach. To facilitate the later presentation, we introduce and
formalize the notion of blended traces.

\begin{definition}(Blended Trace)
	Given a symbolic trace $\sigma$ and multiple concrete traces,
        $\epsilon_{1},\cdot\cdot\cdot,\epsilon_{n}$ that traverse the
        same program path as $\sigma$, a blended trace, $\lambda$, is a
        sequence of the form
        $(\theta_{i}\rightarrow\theta_{i+1})^*$, where $\theta_{i}$ is
        an ordered pair <$e_{i},S_{i}$>, where $e_{i}$ is a statement
        in $\sigma$ and $S_{i}= \{s_{i\_1},\cdot\cdot\cdot,s_{i\_{N_{\epsilon}}}\}$
        is the set of program states $e_{i}$ created in
        $\epsilon_{1},\cdot\cdot\cdot,\epsilon_{N_{\epsilon}}$.
\end{definition}

For simplicity, we assume that a blended trace $\lambda_{i}$ is
composed of $\sigma_{i}$ and two concrete traces, $\epsilon_{i\_1}$
and $\epsilon_{i\_2}$.  Given the $j$-th ordered pair in
$\lambda_{i}$, the first RNN computes the embedding of the statement
$(x_{i\_j\_w_{s_{1}}},\cdot\cdot\cdot,x_{i\_j\_w_{s_{N_{s}}}})$ to be
(based on an abstraction of the computation defined in
Equation~\ref{equ:1}):
\begin{equation}
h_{i\_j\_w_{s_{N_{s}}}} = f_{1}(x_{i\_j\_w_{s_{N_{s}}}},h_{i\_j\_w_{s_{N_{s}-1}}}) \notag \label{equ:11}
\end{equation}
The second RNN embeds the two program states in the $j$-th ordered pair as
\begin{align}
h_{i\_1\_j\_v_{N_{v}}} &= f_{2}(x_{i\_1\_j\_v_{N_{v}}},h_{i\_1\_j\_v_{N_{v}-1}}) \notag\\ 
h_{i\_2\_j\_v_{N_{v}}} &= f_{2}(x_{i\_2\_j\_v_{N_{v}}},h_{i\_2\_j\_v_{N_{v}-1}}) \notag \label{equ:122}
\end{align}

Now we present the idea key to \liger's success. To combine the
strengths of both approaches, we fuse the vector representations
across the feature dimensions to compute a single embedding of each
ordered pair in a blended trace. Specifically,
we adopt the attention mechanism to allocate a weight 
for each feature vector. Assume $H^{e}_{i\_j-1}$ denotes the embedding 
that represents $\lambda_{i}$ before $j$-th ordered pair\footnote{We give the 
formal definition of $H^{e}_{i\_j-1}$ in the next paragraph.}, we 
compute the attention weights of $h_{i\_j\_w_{s_{N_{s}}}}$, 
$h_{i\_1\_j\_v_{N_{v}}}$ and $h_{i\_2\_j\_v_{N_{v}}}$ below:
\begin{align}
\alpha_{i\_j\_w_{s_{N_{s}}}} &= \frac{\mathit{exp}(\mu_{i\_j\_w_{s_{N_{s}}}})}{\mathit{exp}(\mu_{i\_j\_w_{s_{N_{s}}}})+\mathit{exp}(\mu_{i\_1\_j\_v_{N_{v}}})+\mathit{exp}(\mu_{i\_2\_j\_v_{N_{v}}})} \notag\\
\alpha_{i\_1\_j\_v_{N_{v}}} &= \frac{\mathit{exp}(\mu_{i\_1\_j\_v_{N_{v}}})}{\mathit{exp}(\mu_{i\_j\_w_{s_{N_{s}}}})+\mathit{exp}(\mu_{i\_1\_j\_v_{N_{v}}})+\mathit{exp}(\mu_{i\_2\_j\_v_{N_{v}}})} \notag\\
\alpha_{i\_2\_j\_v_{N_{v}}} &= \frac{\mathit{exp}(\mu_{i\_2\_j\_v_{N_{v}}})}{\mathit{exp}(\mu_{i\_j\_w_{s_{N_{s}}}})+\mathit{exp}(\mu_{i\_1\_j\_v_{N_{v}}})+\mathit{exp}(\mu_{i\_2\_j\_v_{N_{v}}})} \notag
\end{align}
where $\oplus$ denotes vector concatenation and
$\mu_{i\_j\_w_{s_{N_{s}}}}$, $\mu_{i\_1\_j\_v_{N_{v}}}$ and
$\mu_{i\_2\_j\_v_{N_{v}}}$ are defined as follows ($a_{1}$ stands for a
feedforward neural network):
\begin{align}
\mu_{i\_j\_w_{s_{N_{s}}}} &= a_{1}(h_{i\_j\_w_{s_{N_{s}}}} \oplus H^{e}_{i\_j-1})    \notag\\
\mu_{i\_1\_j\_v_{N_{v}}} &= a_{1}(h_{i\_1\_j\_v_{N_{v}}} \oplus H^{e}_{i\_j-1})  \notag\\
\mu_{i\_2\_j\_v_{N_{v}}} &= a_{1}(h_{i\_2\_j\_v_{N_{v}}} \oplus H^{e}_{i\_j-1})  \notag
\end{align}
Using the attention weights, we compute the embedding of the $j$-th
ordered pair in $\lambda_{i}$ as
\begin{equation}
h_{i\_j} = \alpha_{i\_j\_w_{s_{N_{s}}}} * h_{i\_j\_w_{s_{N_{s}}}} + \alpha_{i\_1\_j\_v_{N_{v}}} * h_{i\_1\_j\_v_{N_{v}}} + \alpha_{i\_2\_j\_v_{N_{v}}} * h_{i\_2\_j\_v_{N_{v}}} \notag
\end{equation}
Note that we evenly distribute the weights across all feature vectors
embedded from the first ordered pair in any blended trace.

\vspace*{3pt}
\noindent
{\textbf{Executions Embedding Layer.}}
Given $h_{i\_1},\cdot\cdot\cdot,h_{i\_|\lambda_{i}|}$, the embeddings for
all ordered pairs in $\lambda_{i}$, we use the third RNN to model the
flow of the blended trace.
\begin{equation}
H^{e}_{i} = f_{3}(H^{e}_{i\_|\lambda_{i}|-1},h_{i\_|\lambda_{i}|}) \notag
\end{equation}
where $H^{e}_{i\_j}$ is the embedding that represents the partial
blended trace from the first ordered pair to the $j$-th ordered pair
(including the $j$-th ordered pair). In other words, $H^{e}_{i}$
represents the entire $\lambda_{i}$.

\vspace*{3pt}
\noindent
{\textbf{Programs Embedding Layer.}}
We design a pooling layer to compress the embeddings of all the blended
traces, $H^{e}_{1},\cdot\cdot\cdot,H^{e}_{U}$, one for each path to a
program embedding $\mathcal{H}_{P}$.
\begin{equation}
\mathcal{H}_{P} = \mathit{max\_pooling}(H^{e}_{1},\cdot\cdot\cdot,H^{e}_{U}) \notag
\end{equation}

\vspace*{3pt}
\noindent
{\textbf{Loss Function.}}
The network is trained to minimize the cross-entropy loss on a softmax
over the prediction labels. Training is performed using a gradient
descent based algorithm which utilizes back-propagation to adjust the
weight of each parameter.

\begin{figure*}[t]
	\vskip 0.2in
	\begin{center}
	  \centerline{\includegraphics[width=.9\columnwidth]{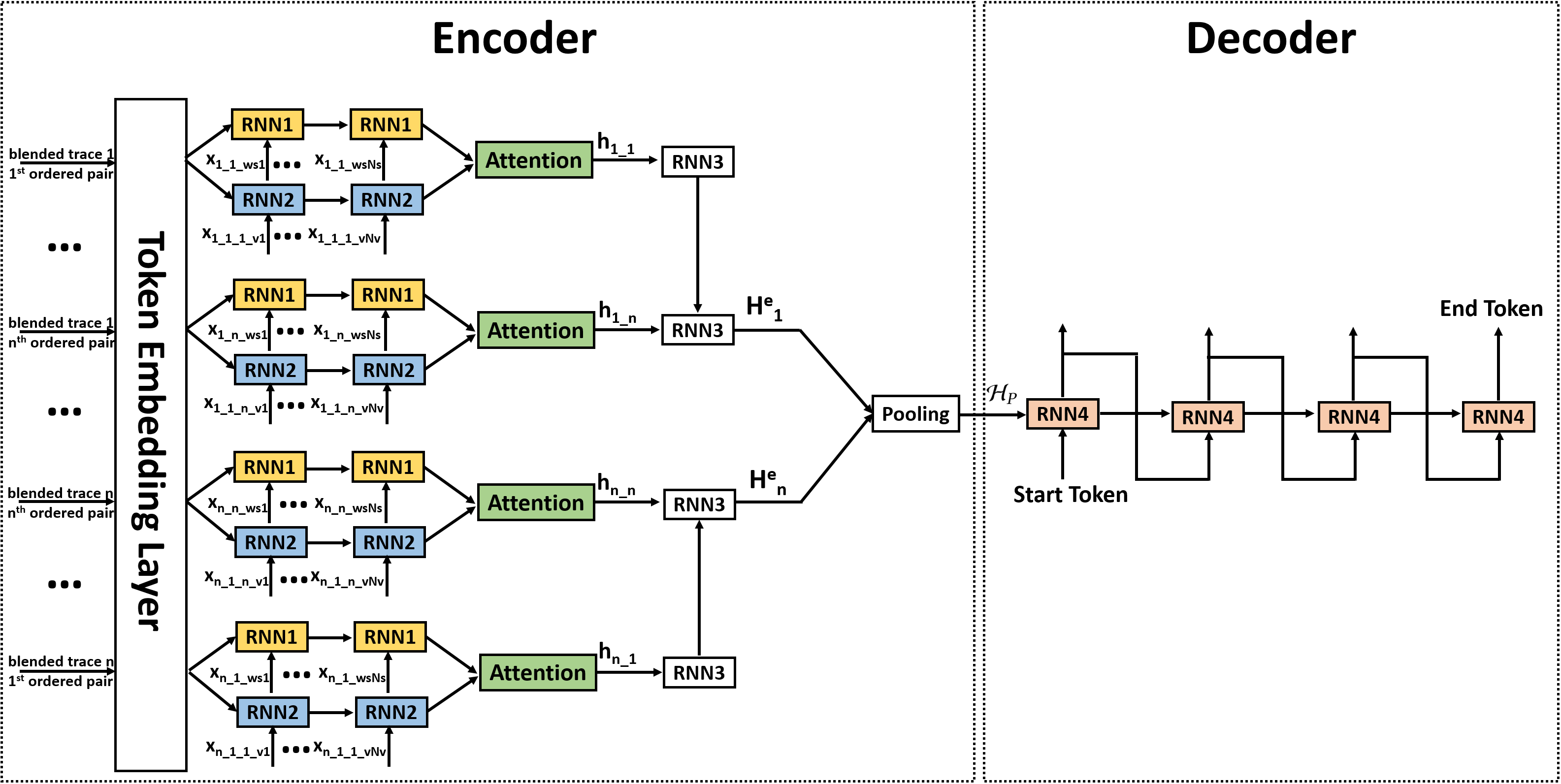}}
          \vspace*{-3pt}
	  \caption{Extending \liger into an encoder-decoder architecture.}
	  \label{fig:ende}
	\end{center}
	\vskip -0.3in
\end{figure*}

\subsection{Extension of \liger}

We also extend \liger into an encoder-decoder architecture to solve
the problem of method name prediction. Specifically, we remove the
program embedding layer from \liger and add a decoder to predict
method names as sequences of words.  Figure~\ref{fig:ende} depicts the
extended architecture.

\vspace*{3pt}
\noindent
{\textbf{Encoder.}}
We cut \liger's last layer to play the role of the encoder. Given a
program $P$, the encoder outputs $\mathcal{H}_{P}$, and
$\{\{H^{e}_{i\_j}|j\in[1,|\lambda_{i}|]\}|i\in[1,U]\}$, the set of
embeddings for each blended trace of program $P$.

\vspace*{3pt}
\noindent
{\textbf{Decoder.}}
We use another RNN to be the decoder. For initialization, we provide
the decoder with the program embedding $\mathcal{H}_{P}$. The decoder
also receives a special token to begin, and emits another to end the
generation.

\vspace*{3pt}
\noindent
{\textbf{Attention.}}
As explained in Section~\ref{subsec:atte}, we incorporate the
attention mechanism to the extended architecture to aid the decoding
process. Unlike the attention neural network introduced previously
where the decoder attends over the input symbols from a single source
sentence, our decoder attends over the flow of all blended traces (\ie
$\{\{H^{e}_{i\_j}|j\in[1,|\lambda_{i}|]\}|i\in[1,U]\}$).

We recompute the context vector $c_{t}$ (defined in
Equation~\ref{equ:6}) for each generated word $y_{t}$ as:
\begin{equation}
c_{t} = \sum_{i=1}^{U} \sum_{j=1}^{|\lambda_{i}|} \alpha_{t\_i\_j}H^{e}_{i\_j}     \notag
\end{equation}
Each attention weight $\alpha_{t\_i\_j}$ is computed by
\begin{equation}
\alpha_{t\_i\_j} = \frac{\mathit{exp}(\mu_{\mathit{t\_i\_j}})}{\sum_{i=1}^{U}\sum_{j=1}^{|\lambda_{i}|}\mathit{exp}(\mu_{t\_i\_j})}   \notag   
\end{equation}
where
$\mu_{t\_i\_j} = a_{2}(H^{d}_{t-1},H^{e}_{i\_j})$ 
is the attention score which measures how well each $H^{e}_{i\_j}$
correlates with the previous hidden state of the decoder,
$H^{d}_{t-1}$. The parameter $a_{2}$ stands for a feedforward neural
network that is jointly trained with other components in the extended
architecture. Figure~\ref{fig:attention} depicts a graphical
illustration of the attention mechanism built into the extended
architecture.

%% file: impl.tex
\section{Implementation}

In this section, we first describe how to parse programs into \liger's
required format, and then present the implementation details of \liger.

\subsection{Data Pre-Processing}

To obtain the data format \liger requires, we weigh the option of
running a symbolic execution engine to obtain symbolic
traces. However, as \liger does not particularly require executions to
achieve high code coverage and the practical limitations of symbolic
execution complicates our implementation, we choose to execute
programs with random inputs to obtain the concrete traces, from which
we derive the symbolic traces. As a final step, we construct the
blended traces.

Since we mostly deal with C\# and Python programs, we use the
Microsoft Roslyn compiler framework and IronPython for those program
pre-processing tasks, including extracting a program's abstract syntax
tree (AST) and instrumenting the program's source code. As for program
executions, we use Roslyn's emit API and IronPython's interpreter.

\subsection{Implementation of \liger}

\liger is implemented in Tensorflow. All RNNs have one single
recurrent layer with 100 hidden units. We adopt random initialization
for weight initialization. Our vocabulary has 7,188 unique tokens (\ie,
tokens of all programs and the values of all variables at each time
step), each of which is embedded into a 100-dimensional vector. All
networks are trained using the Adam optimizer~\cite{kingma2014adam}
with the learning and the decay rates set to their default values
(\ie, learning rate = 0.0001, beta1 = 0.9, beta2 = 0.999) and a mini-batch
size of 100.  We use five Red Hat Linux servers, each of which hosts 
four Tesla V100 GPUs (with 32GB GPU memory).

\begin{figure*}[!tbp]
	\vskip 0.2in
	\begin{center}
		\centerline{\includegraphics[width=.7\columnwidth]{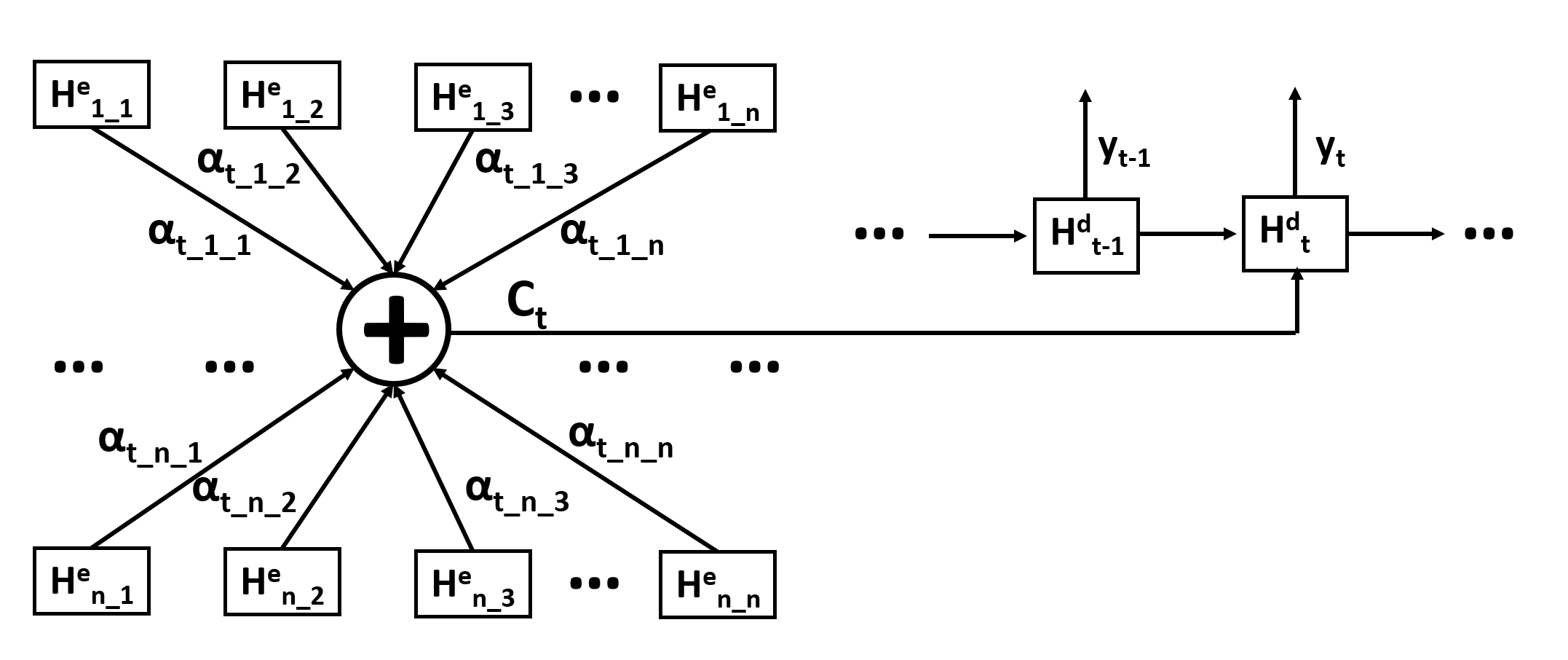}}
		\caption{Graphic illustration of the attention
                  mechanism in the extended architecture.}
		\label{fig:attention}
	\end{center}
	\vskip -0.2in
\end{figure*}

%% file: eva.tex
\section{Evaluation}
\label{sec:eva}

This section presents the details of our extensive and comprehensive
evaluation. First, we evaluate \liger on \coset.  Then, we perform 
ablations to study and evaluate \liger's internal design and
realization. Finally, we evaluate how well the extension of \liger
predicts method names.

\subsection{Evaluation of \liger on \coset}

We briefly introduce \coset and then present in detail each of the
conducted experiments.

\vspace{0.2cm}
\subsubsection{Introduction of \coset}~\\[3pt]
\coset is a recently proposed benchmark framework that aims at
providing a consistent baseline for evaluating the performance of neural
program embeddings.

\vspace*{3pt}
\noindent
{\textbf{\coset's Dataset.}}  \coset contains a dataset consisting of
close to 85K programs developed by a large number of programmers while
solving ten coding problems. Programs are handpicked to ensure
sufficient variations (\eg, coding style, algorithmic choices or
structure). The dataset was manually analyzed and labeled. The work
was done by fourteen PhD students and exchange scholars at the
University of California, Davis. To reduce the labeling error, most of
the programs solving the same coding problem are assigned to a single
participant. Later the results are cross-checked among all
participants for validation.  The whole process took more than three
months to complete.  Participants came from different research
backgrounds such as programming language, database, security,
graphics, machine learning, \etc All participants were interviewed and
tested for their knowledge of program semantics. The labeling is on
the basis of execution semantics (\eg, bubble sort, insertion sort,
merge sort, \etc, for a sorting routine).  Certain variations are
discounted to keep the total number of labels manageable. For example,
local variables for temporary storage are ignored, so are iterative
styles (looping or recursion) and the sorting order (descending or
ascending). After obtaining \coset, we removed the problem of printing
the chessboard whose execution does not require external inputs.  Readers
are invited to consult the supplemental materials for the descriptions
of all coding problems and the list of all labels.

To collect execution traces, we run each program with a large number
of randomly generated inputs. Our intention is to trigger a large
fraction of the program paths, and cover each path with a sufficient
amount of executions.\footnote{We generate 18 per program, each of
  which is covered by 5 executions, to fully utilize a V100 GPU.} We
remove programs that fail to pass all test cases (\ie, crashes or
having incorrect results) from the dataset. In the end, we are left
with 65,596 programs which we split into a training set containing
45,596 programs, a validation set of 9,000 programs and a test set with
the remaining 9,000 programs (Table~\ref{Table:data}).

\begin{table}[htbp!]
	\begin{center}
		\begin{adjustbox}{max width=.9\textwidth}
			\begin{tabular}{c | c | c | c } 
				\hline
				\textbf{Benchmarks} 
				& \textbf{Training}
				& \textbf{Validation}
				& \textbf{Testing} \\
				\hline		
				Find Array Max Difference &5,821  &1,000 &1,000 \\ 
				\hline
				Check Matching Parenthesis  &3,451 &1,000  &1,000 \\ 
				\hline
				Reverse a String  &6,087  &1,000  &1,000 \\ 
				\hline		
				Sum of Two Numbers  &6,635  &1,000  &1,000 \\							
				\hline		
				Find Extra Character  &5,056  &1,000  &1,000 \\ 			
				\hline		
				Maximal Square  &4,862  &1,000  &1,000 \\ 			
				\hline		
				Maximal Product Subarray &5,217  &1,000  &1,000 \\ 				
				\hline		
				Longest Palindrome &5,480  &1,000  &1,000 \\ 				
				\hline		
				Trapping Rain Water &2,987  &1,000  &1,000 \\							
				\hline
				\hline
				Total &45,596 &9,000  &9,000 \\ 
				\hline
			\end{tabular}
		\end{adjustbox}
	\end{center}
	\caption{Dataset used in semantic classification.}
	\label{Table:data}
\end{table}

\vspace*{3pt}
\noindent
{\textbf{\coset's Task.}}  The prediction task that \coset provides is
semantic classification. That is a model is required to predict the
category where a program falls based on its semantics. In other words,
not only does the model need to classify which coding problem that a
program solves, but also which strategy that the program adopts in its
solution.  We use prediction accuracy and F1 score as the evaluation
metrics.

\vspace*{3pt}
\noindent
{\textbf{\coset's Transformations.}}  \coset also includes a suite of
program transformations. These transformations when applied to the
base dataset can simulate natural changes to program code due to
optimization and refactoring. Wang~\etal use those transformations in
\coset for two purposes. First, they are used to measure the stability
of model predictions.  By applying only semantically-preserving
transformations, one can test how frequently a model changes its
original predictions. The more frequently a prediction changes, the
less precise the model is. Second, when a model makes an incorrect
classification, the transformations are used to identify the root
cause of a misclassification for debugging. In this paper, we measure
the stability of each model by applying semantically-preserving
transformations on the \coset's dataset.

\vspace{0.2cm}
\subsubsection{Evaluating \liger's performance}~\\
\label{subsubsec:cos}
This experiment compares \liger with several other DNNs targeting
learning program representations.
\begin{itemize}
	\item \textbf{GGNN}: GGNN is first proposed
          by~\citet{li2015gated} and later utilized
          by~\citet{allamanis2017learning} to predict variable misuse
          bugs in a program. The idea is to construct a graph out of a
          program's AST along with additional semantic edges that are
          manually designed to improve a graph's expressiveness. Those
          edges denote variable read/write relations, orders of leaf
          nodes in the AST, guarding conditions of statements, \etc In
          this experiment, we strip off the prediction layer and
          average all nodes in a graph to be the program embedding.
		
	\item \textbf{code2vec}: code2vec is proposed
          by~\citet{Alon:2019:CLD:3302515.3290353} to predict method
          names based on a large, cross-project corpus. The approach
          first decomposes the code for a program to a collection of
          paths in its abstract syntax tree, and then learns to
          represent each path as well as aggregating a set of them. We
          use code vector --- the aggregation of context vectors ---
          to represent a program.
	
	\item \textbf{\dypro}: We also use
          \dypro~\cite{wang2019learning}, another DNN that learns
          program representations from program executions. To allow a
          meaningful direct comparison, we feed \liger and \dypro the
          same collection of execution traces and observe differences
          in their performance.
\end{itemize}

\begin{figure*}[htbp!]
	\begin{subfigure}[b]{0.425\textwidth}
		\begin{center}
			\centerline{\includegraphics[width=\columnwidth]{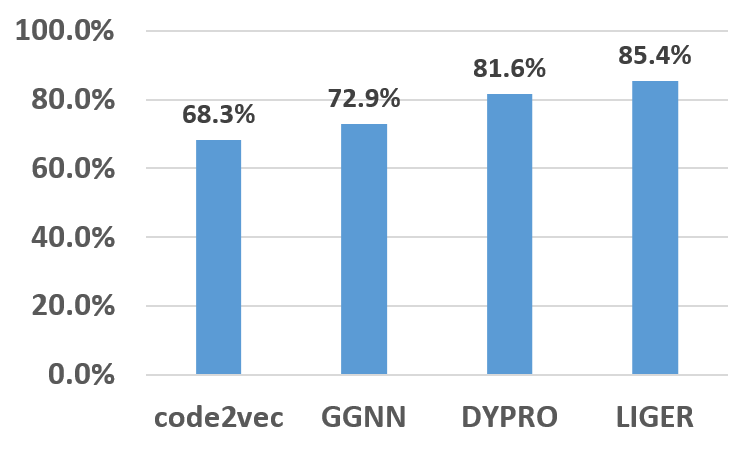}}
			\caption{The comparison of models using accuracy.}
			\label{fig:accuracy}
		\end{center}
	\end{subfigure}
	\;
	\begin{subfigure}[b]{0.425\textwidth}
		\begin{center}
			\centerline{\includegraphics[width=\columnwidth]{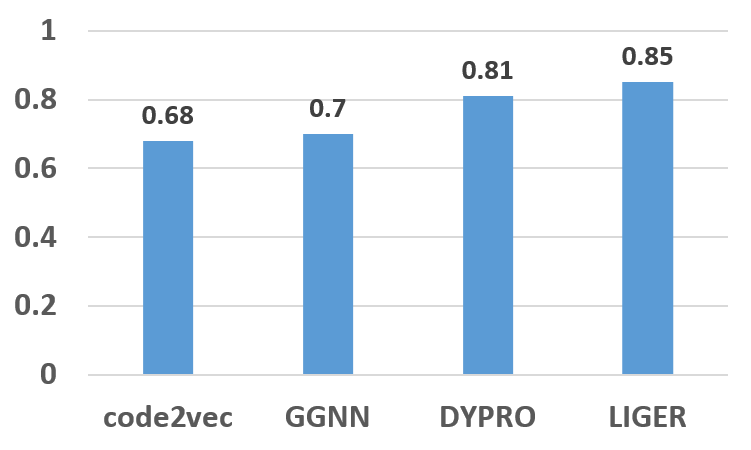}}
			\caption{The comparison of models using F1 score.}
			\label{fig:accuracyF1}
		\end{center}
	\end{subfigure}	
	
	\caption{Comparing all models with the semantic classification task.}
\end{figure*}

As depicted in Figure~\ref{fig:accuracy}, \liger is the most accurate
model among all, albeit not a substantial improvement over \dypro. In
contrast, static models (both GGNN and code2vec) perform significantly
worse. In terms of the F1 score, \liger also achieves the best results
and significantly outperforms GGNN and code2vec
(Figure~\ref{fig:accuracyF1}).

\begin{figure*}[htb!]
	\begin{subfigure}[b]{0.425\textwidth}
		\begin{center}
			\centerline{\includegraphics[width=\columnwidth]{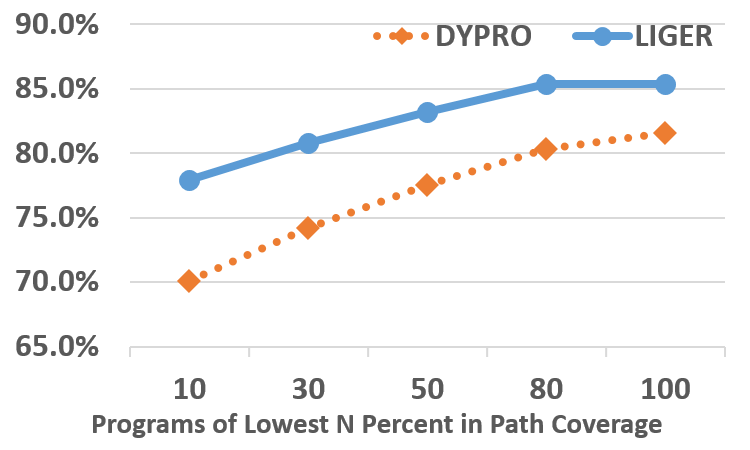}}
			\caption{Comparison on prediction accuracy. }
			\label{fig:lipro}
		\end{center}
	\end{subfigure}
	\;
	\begin{subfigure}[b]{0.425\textwidth}
		\begin{center}
			\centerline{\includegraphics[width=\columnwidth]{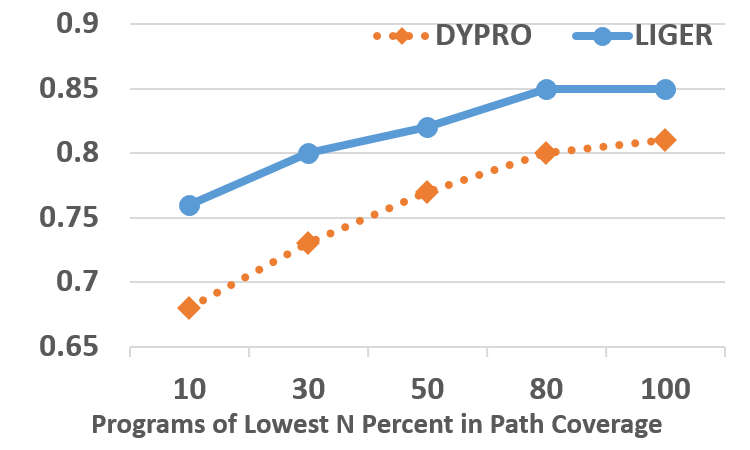}}
			\caption{Comparison on F1 score. }
			\label{fig:liproF1}
		\end{center}
	\end{subfigure}	
	
	\caption{Comparison of \liger against \dypro for programs with
          increasing path coverage.}
	\label{fig:liproT}
\end{figure*}

We performed another experiment for a more in-depth understanding of
the comparison between \liger and \dypro. In particular, we split
\coset's testing programs into subgroups according to their path
coverage (\ie, lowest 10\% of programs to 100\% in terms of path
coverage) and evaluate how the models compare on each subgroup. We
reuse both models from the previous experiment without retraining.  As
depicted in Figures~\ref{fig:lipro} and~\ref{fig:liproF1}, when tested
on the lowest 30\% of programs ranked by path coverage, \liger
outperforms \dypro by a wide margin. The gap shrinks on programs with
higher coverage, indicating similar performances of the two models on
these programs. The results indicate that, although \liger does not
significantly improve \dypro in overall accuracy, it is more reliable
to generalize to unseen programs, particularly those on which high
path coverage is difficult to obtain. Subsequent experiments are
designed to understand the reasons behind this observation. 

\vspace{0.2cm}
\subsubsection{Reliance on Program Executions}~\\
\label{subsubsec:reli}
We examine to what degree \liger relies on executions to produce
precise program embeddings. Reusing the semantic classification task,
we evaluate \liger from two aspects. First, we randomly reduce the
number of concrete traces used to construct a blended trace while
keeping the total number of symbolic traces constant for each program
in both \coset's training and testing sets.  In other words, we aim to
find out how \liger would perform when each symbolic trace is
accompanied with a fewer number of concrete traces.

\begin{figure*}[t!]
	\begin{subfigure}[b]{0.425\textwidth}
		\begin{center}
			\centerline{\includegraphics[width=\columnwidth]{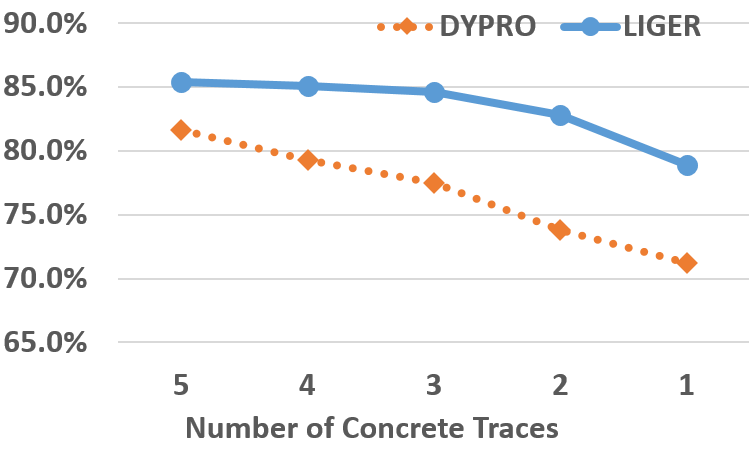}}
			\caption{Change of accuracy for \liger when number of executions per path is randomly reduced.}
			\label{fig:reducepro}
		\end{center}
	\end{subfigure}
	\;
	\begin{subfigure}[b]{0.425\textwidth}
		\begin{center}
			\centerline{\includegraphics[width=\columnwidth]{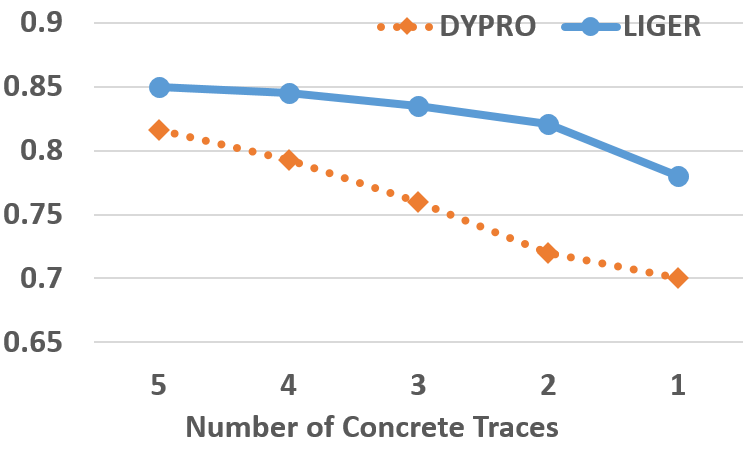}}
			\caption{Change of F1 score for \liger when
                          the number of executions per path is randomly
                          reduced.}
			\label{fig:reduceproF1}
		\end{center}
	\end{subfigure}	
	
	\caption{\liger's results when the number of executions per
          path is randomly reduced.}

\end{figure*}

Figures~\ref{fig:reducepro} and~\ref{fig:reduceproF1} show results for
this experiment. In general, reducing concrete traces in a blended
trace path has a small effect on \liger's performance. Perhaps more
unexpectedly, \liger exhibits almost the same accuracy and F1 score
when it is supplied with no less than three concrete traces. To dig
deeper, we also investigate how attention weights of each symbolic
trace change when executions are down-sampled. We observe that, upon
model convergence, the attention weight for each statement along each
symbolic trace is 5.98 on average, and the weight stays largely
unchanged throughout the reduction.  Furthermore, the rest of the
attention weight (\ie, 4.12) is almost evenly split into the concrete
traces regardless their number.  This finding shows that \liger relies
more on the static (symbolic) feature dimension while generalizing
from the training programs. Meanwhile, it views concrete traces as
parallel instantiations of the same symbolic trace. Most importantly,
\liger is capable of compensating the loss of concrete traces by
increasing the importance of the remaining ones, therefore keeping its
accuracy constant.  Specifically, when left with two concrete traces
in a blended trace, \liger can still achieve more than a 83\% (\resp
0.82) accuracy (\resp F1 score).  In contrast, \dypro suffers a
significant performance drop, empirically confirming its far higher
demand for concrete traces.

\begin{figure*}[htbp!]
	\begin{subfigure}[b]{0.425\textwidth}
		\begin{center}
			\centerline{\includegraphics[width=\columnwidth]{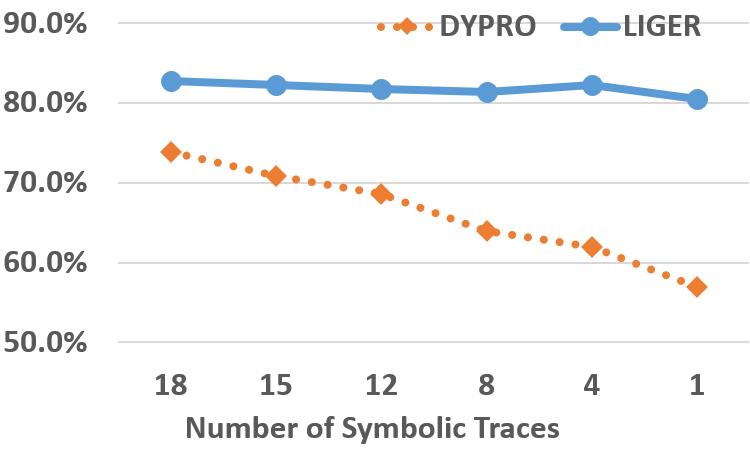}}
			\caption{Change of accuracy for \dypro and
                          \liger when program paths are reduced but
                          branch coverage is preserved for each
                          program.}
			\label{fig:reducePath}
		\end{center}
	\end{subfigure}
	\;
	\begin{subfigure}[b]{0.425\textwidth}
		\begin{center}
			\centerline{\includegraphics[width=\columnwidth]{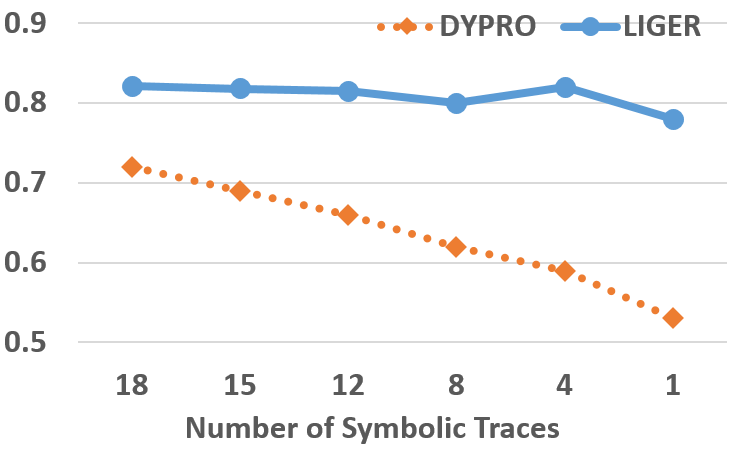}}
			\caption{Change of F1 score for \dypro and
                          \liger when program paths are reduced but
                          branch coverage is preserved for each
                          program.}
			\label{fig:reducePathF1}
		\end{center}
	\end{subfigure}	
	
	\caption{Accuracy trend for \liger and \dypro when branch
          coverage is preserved for each program throughout path
          reduction.}
	\label{fig:comp}
\end{figure*}

Next, we investigate how \liger reacts when the number of symbolic
traces decreases.  We have identified a minimum set of symbolic traces
for each program in \coset's dataset that achieve the same branch
coverage as before.\footnote{We apply a greedy heuristic to pick the
  symbolic trace that covers the most uncovered branches.} We remove
symbolic traces that are not in the minimum set and
examine how path coverage affects \liger's accuracy. 
In this experiment, we randomly select two out of the original 
five concrete traces, from which we generate a minimum set of 
blended traces. 
As a baseline, we show how \liger compares against \dypro on the same
concrete traces throughout path reduction.

When branch coverage is preserved during path reduction, 
\liger's performance is largely unaffected (Figure~\ref{fig:comp}), 
indicating its strong resilience to reduced program 
paths.
%
%
%
Even if training and testing on the minimum set of blended traces,
\liger is more accurate than \dypro trained and tested on the entire
set of concrete traces (82.3\% vs. 81.6\% in accuracy, and 0.82 vs.
0.81 in F1 score). As the average size of the minimum set of symbolic
traces is calculated to be 4.7 (\ie, 9.4 concrete executions) for each
program, \liger used nearly 10x fewer executions covering almost
74\% fewer program paths. By learning from the minimum set of blended
traces, \liger also reduces the training time from 273 hours to 38
hours under the same setup.

Our findings indicate that \liger depends far less on program
executions than \dypro. 
In addition, the results also explain the superior performance \liger
exhibits on the \coset's testing programs that are difficult to cover
(Figure~\ref{fig:liproT}).  Because \liger does not require very high
code coverage in order to produce precise program embeddings, it
lessens its reliance on both the number and diversity of program
executions.

\begin{figure*}[t!]
	\begin{subfigure}[b]{0.425\textwidth}
		\begin{center}
			\centerline{\includegraphics[width=\columnwidth]{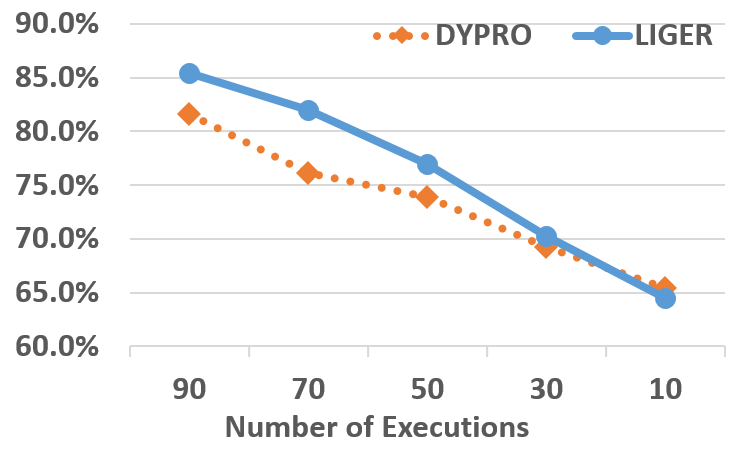}}
			\caption{Change of accuracy for \dypro and
                          \liger when executions are randomly
                          reduced.}
			\label{fig:accuracyChange}
		\end{center}
	\end{subfigure}
	\;
	\begin{subfigure}[b]{0.425\textwidth}
		\begin{center}
			\centerline{\includegraphics[width=\columnwidth]{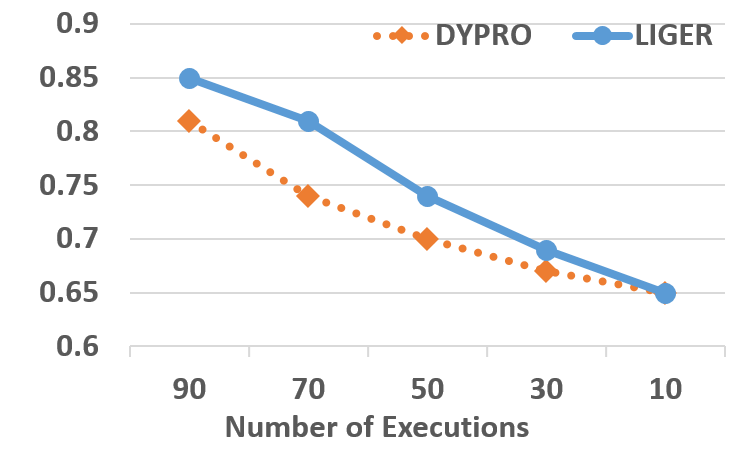}}
			\caption{Change of F1 score for \dypro and
                          \liger when executions are randomly
                          reduced.}
			\label{fig:accuracyChangeF1}
		\end{center}
	\end{subfigure}	
	
	\caption{Accuracy trend for \liger and \dypro when randomly
          down-sampling executions.}
	\label{fig:accuracyChangeT}
\end{figure*}

To verify the validity of these two experiments, we conduct another
study in which we randomly reduce both symbolic and concrete traces,
and construct blended traces accordingly. Results
(Figure~\ref{fig:accuracyChangeT}) show that \liger overall still
performs slightly better than \dypro.  However, it is not nearly as
tolerant of data loss as the two above experiments have shown.  Thus,
to conclude, compared to \dypro, \liger displays much less reliance on
program executions. Specifically, we show that neither the reduction
of blended traces nor the decrease in path coverage 
significantly affects \liger's performance.


\vspace{0.2cm}
\subsubsection{Comparison on Model Stability}~\\
In this experiment, we measure the stability for each model prediction
by applying the semantically-preserving program transformations to
\coset's test set. Those transformations are routinely performed by
compilers for code optimization (\eg, constant and variable
propagation, dead code elimination, loop unrolling and
hoisting). Specifically, we apply the transformations to each test
program in Table~\ref{Table:data} to create a new test set. Programs
with no applicable transformations are excluded from the new test
set. We then examine if the models make the same prior predictions in
the semantic classification task.

Table~\ref{Table:stability} presents the results. The number in each
cell denotes the percentage of programs in the new test set on which a
model has changed its prediction. Overall, \liger is the most stable
model against those program transformations, and is significantly more
stable than both static models.

\begin{table}[htbp!]
	\begin{center}
		\begin{adjustbox}{max width=.9\textwidth}
			\begin{tabular}{c | c | c | c | c }
				\hline
				\textbf{Model} 				
				& \begin{tabular}{@{}c@{}} \textbf{Constant and Variable}  \\ \textbf{Propagation} \end{tabular}
				& \begin{tabular}{@{}c@{}} \textbf{Dead Code}  \\ \textbf{Elimination} \end{tabular}
				& \textbf{Loop Unrolling}
				& \textbf{Hoisting} \\
				\hline		
				code2vec &8.6\%  &12.4\%  &23.9\%  &12.2\% \\ 
				\hline
				GGNN &6.8\% &7.4\%  &16.8\% &10.9\% \\ 
				\hline
				\dypro &3.4\% &4.1\%  &6.8\% &1.8\% \\ 
				\hline					
				\textbf{\liger} &\textbf{3.4\%} &\textbf{3.5\%}  &\textbf{7.9\%} &\textbf{1.3\%} \\ 
				\hline					
			\end{tabular}
		\end{adjustbox}
	\end{center}
	\caption{Results on measuring model stability, measured as
          percentage of changed predictions when applying
          semantics-preserving transformations.}
	\label{Table:stability}
\end{table}

\begin{figure*}[t!]
	\begin{subfigure}[b]{0.425\textwidth}
		\begin{center}
			\centerline{\includegraphics[width=\columnwidth]{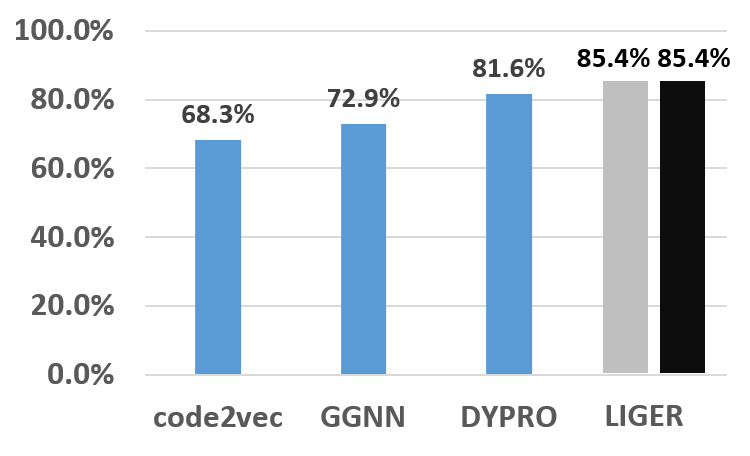}}
			\caption{Change of accuracy before and after
                          removing the static feature dimension from
                          \liger.}
			\label{fig:abasta}
		\end{center}
	\end{subfigure}
	\;
	\begin{subfigure}[b]{0.425\textwidth}
		\begin{center}
			\centerline{\includegraphics[width=\columnwidth]{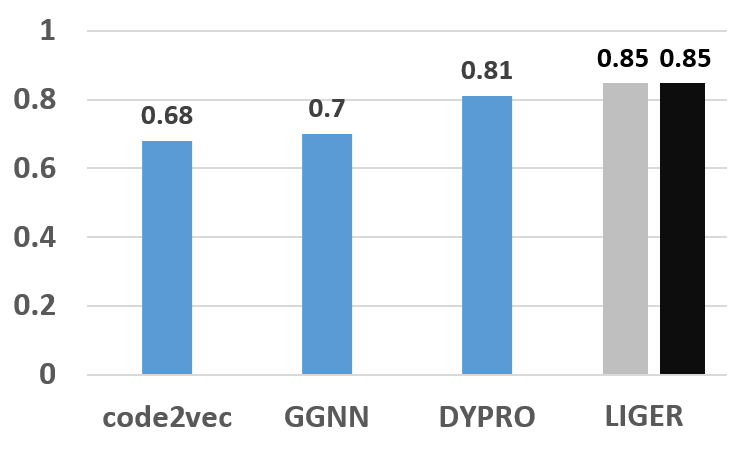}}
			\caption{Change of F1 score before and after
                          removing the static feature dimension from
                          \liger.}
			\label{fig:abastaF1}
		\end{center}
	\end{subfigure}	
	
	\caption{Effects of the static feature dimension in
          \liger for the semantic classification task. Light bars denote
          results before, while dark bars denote results after the
          ablation.}
	\label{fig:abastaT}
\end{figure*}

\subsection{Ablation Study}

In this section, we conduct an ablation study to understand the
contribution of each component in \liger's architecture. Since all
layers except the fusion layer are essential to \liger's
functionality, our ablation study will only examine the
causality among components in the fusion layer, more precisely, the
effect of both feature dimensions as well as the attention mechanism
that fuses the feature dimensions. To evaluate each new configuration
of \liger, we use model accuracy and data reliance as the two metrics
in the semantics classification task.

\vspace{0.2cm}
\subsubsection{Removing Static Feature Dimension}~\\
First, we remove the symbolic trace from the feature
representation. As a result, RNN1 is no longer needed and removed from
\liger's architecture. Note that the resulting configuration is not
identical to \dypro's architecture where an embedding for each
execution is learned separately.
In contrast, \liger learns an embedding for a multitude of 
executions along the same program path. 
We use the same concrete traces for each program in 
\coset to repeat the experiment on semantics classification. 

As depicted in Figure~\ref{fig:abastaT}, after removing the static
feature dimension, \liger exhibits the same prediction accuracy
(hereinafter light/dark shapes denote \liger's results before/after
the ablation) and F1 score.  This indicates that, when given abundant
concrete traces to learn, \liger is able to generalize from the
dynamic features alone, therefore, symbolic traces becomes
expendable. In addition, even without the static feature dimension,
\liger still significant outperforms both static models in accuracy
and F1 score as shown in Figures~\ref{fig:abasta}
and~\ref{fig:abastaF1}.
 
\begin{figure*}[t!]
	\begin{subfigure}[b]{0.425\textwidth}
		\begin{center}
			\centerline{\includegraphics[width=\columnwidth]{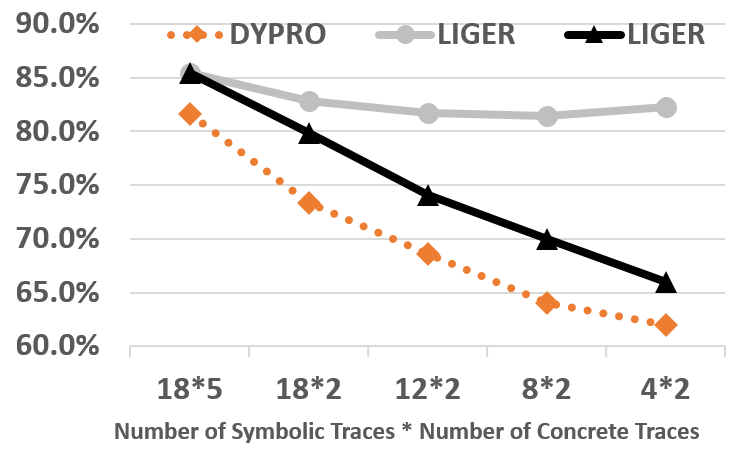}}
			\caption{Comparing \liger's accuracy trend
                          before and after removing static features
                          when both symbolic and concrete traces are
                          reduced. }
			\label{fig:abastachange}
		\end{center}
	\end{subfigure}
	\;
	\begin{subfigure}[b]{0.425\textwidth}
		\begin{center}
			\centerline{\includegraphics[width=\columnwidth]{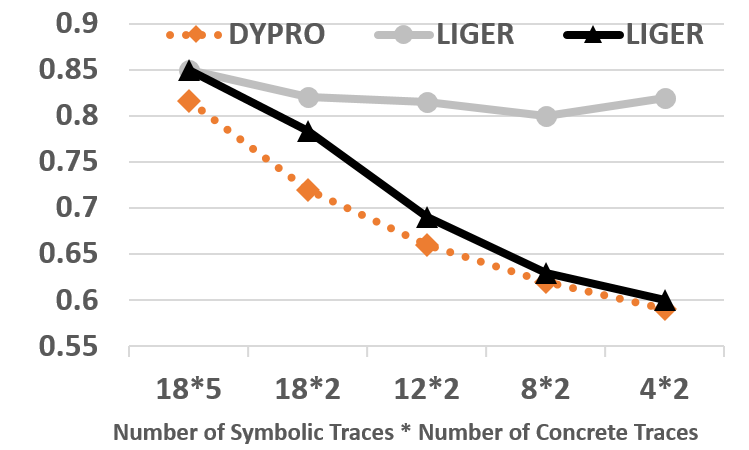}}
			\caption{Comparing \liger's F1 score trend
                          before and after removing static features
                          when both symbolic and concrete traces are
                          reduced.}
			\label{fig:abastachangeF1}
		\end{center}
	\end{subfigure}	
	
	\caption{The effect of static feature dimension on models' 
          reliance on executions.
	}
	\label{fig:abastachangeT}
\end{figure*}

Next, we aim to understand how dependent \liger becomes on executions
after removing the static feature dimension. In our ablation study, we
always maintain the branch coverage of a program while decreasing both
the numbers of symbolic and concrete traces. For brevity, we combine
the results of reducing symbolic and concrete traces in the same
diagram.
As shown in Figures~\ref{fig:abastachangeT}, after removing the
static feature dimension, \liger displays a similar performance trend
to \dypro. In other words, \liger becomes more dependent on program
executions, manifested in the significantly poorer results while
learning from few concrete traces.
This finding reveals that it is the static feature dimension that
contributes to the moderate reliance \liger has on program executions.

\begin{figure*}[hbt!]
	\begin{subfigure}[b]{0.425\textwidth}
		\begin{center}
			\centerline{\includegraphics[width=\columnwidth]{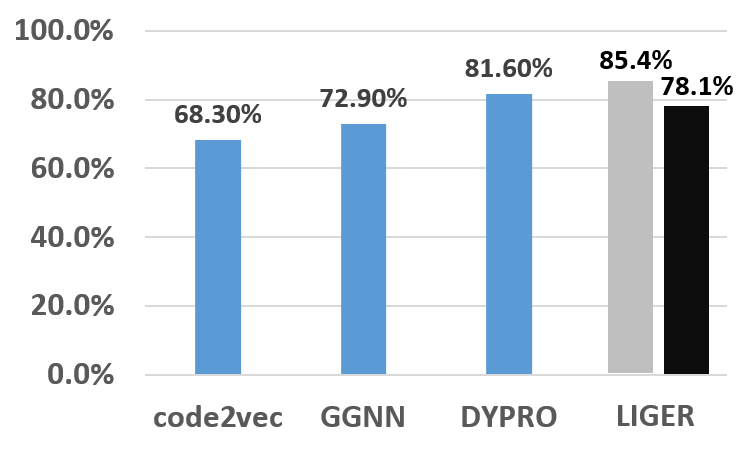}}
			\caption{Change of accuracy before and after
                          removing the dynamic feature dimension from
                          \liger.}
			\label{fig:abadyn}
		\end{center}
	\end{subfigure}
	\;
	\begin{subfigure}[b]{0.425\textwidth}
		\begin{center}
			\centerline{\includegraphics[width=\columnwidth]{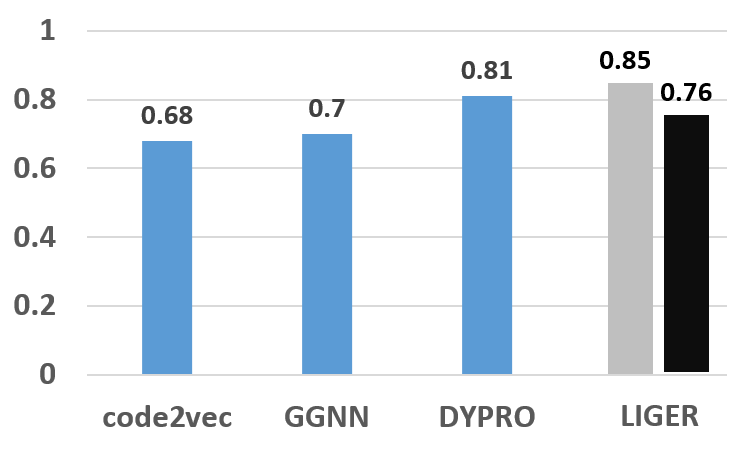}}
			\caption{Change of F1 score before and after
                          removing the dynamic feature dimension from
                          \liger.}
			\label{fig:abadynF1}
		\end{center}
	\end{subfigure}	
	
	\caption{The effect of the dynamic feature dimension in
          \liger for the semantics classification task.}
	\label{fig:abadynT}
\end{figure*}

\vspace{0.2cm}
\subsubsection{Removing Dynamic Feature Dimension}~\\
\label{subsubsec:static}
We remove the dynamic feature dimension from \liger to reveal its
contribution to the entire network. Since \liger is left with the
symbolic traces only, each statement in the trace will receive the
full attention weight in the fusion layer.  Like the prior experiment,
we first measure \liger's accuracy for the semantics classification
task.

As depicted in Figures~\ref{fig:abadyn} and Figure~\ref{fig:abadynF1},
removing dynamic feature has a larger impact on \liger's precision. In
particular, both accuracy and F1 score drop notably despite still
being more accurate than both the static models by a reasonable
margin. This finding confirms the challenges of learning precise
program embeddings from symbolic program features directly. Even
though symbolic traces reflect certain level of the runtime
information, \liger does not manage to learn precise program
embeddings. Next, we study \liger's reliance blended traces.

\begin{figure*}[t!]
	\begin{subfigure}[b]{0.425\textwidth}
		\begin{center}
			\centerline{\includegraphics[width=\columnwidth]{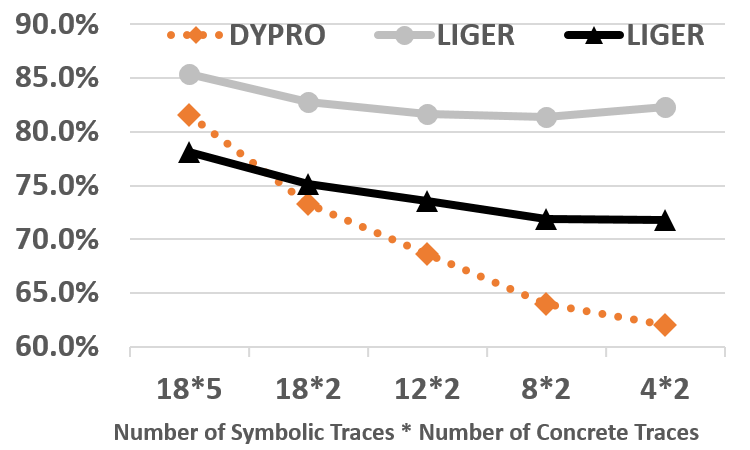}}
			\caption{Comparing \liger's accuracy trend
                          before and after removing the dynamic
                          features both symbolic and concrete traces
                          are reduced. }
			\label{fig:abadynchange}
		\end{center}
	\end{subfigure}
	\;
	\begin{subfigure}[b]{0.425\textwidth}
		\begin{center}
			\centerline{\includegraphics[width=\columnwidth]{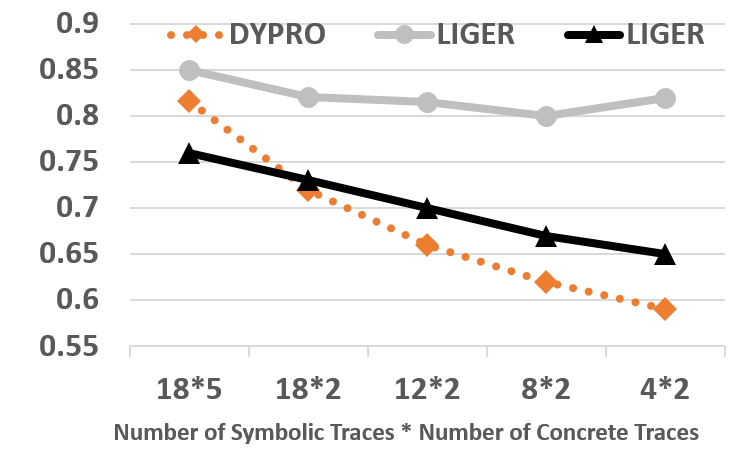}}
			\caption{Comparing \liger's F1 score trend
                          before and after removing the dynamic
                          features both symbolic and concrete traces
                          are reduced.}
			\label{fig:abadynchangeF1}
		\end{center}
	\end{subfigure}	
	
	\caption{The effect of the dynamic feature dimension in \liger
          on the model reliance on executions.
	}
	\label{fig:abadynchangeT}
\end{figure*}

After removing the dynamic feature dimension, \liger is still shown
quite robust against trace reduction. Even though \liger starts at a
lower accuracy and F1 score, it outperforms \dypro as both the numbers of
symbolic and concrete traces decrease
(Figures~\ref{fig:abadynchange} and~\ref{fig:abadynchangeF1}). In
general, the accuracy trend \liger displays correlates well before and
after the removal of dynamic features.  Thanks to the static feature
dimension, \liger does not suffer a significant accuracy drop.

\begin{figure*}[hbt!]
	\begin{subfigure}[b]{0.425\textwidth}
		\begin{center}
			\centerline{\includegraphics[width=\columnwidth]{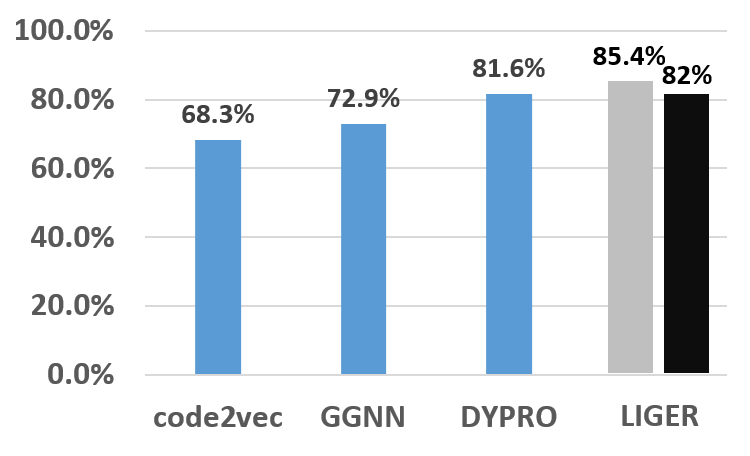}}
			\caption{Change of accuracy before and after
                          removing the attention mechanism from fusion
                          layer.}
			\label{fig:abaatten}
		\end{center}
	\end{subfigure}
	\;
	\begin{subfigure}[b]{0.425\textwidth}
		\begin{center}
			\centerline{\includegraphics[width=\columnwidth]{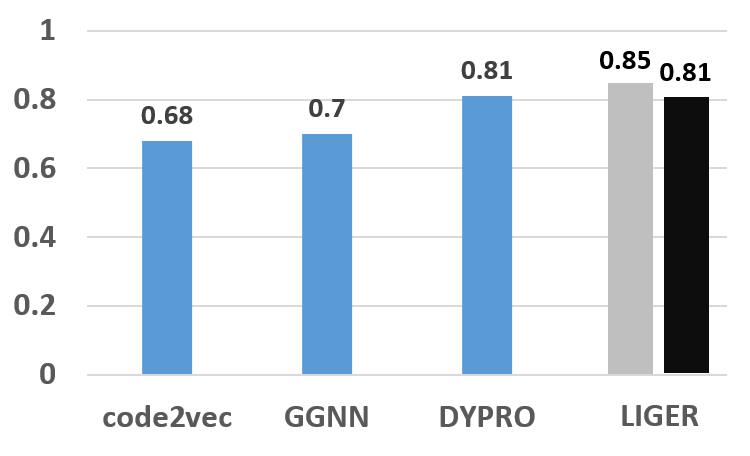}}
			\caption{Change of F1 score before and after
                          removing the attention mechanism from fusion
                          layer.}
			\label{fig:abaattenF1}
			
		\end{center}
	\end{subfigure}	
	
	\caption{The effect of attention in the fusion layer
          for the semantics classification task.}
	\label{fig:abaattenT}
\end{figure*}

\vspace{0.2cm}
\subsubsection{Removing Attention}~\\
Finally, we remove the attention mechanism that controls the fusion of
the two feature dimensions. To keep other components intact in the
fusion layer, we evenly distribute the weights across all traces (\ie,
symbolic and concrete) in a blended trace.

\begin{figure*}[t!]
	\begin{subfigure}[b]{0.425\textwidth}
		\begin{center}
			\centerline{\includegraphics[width=\columnwidth]{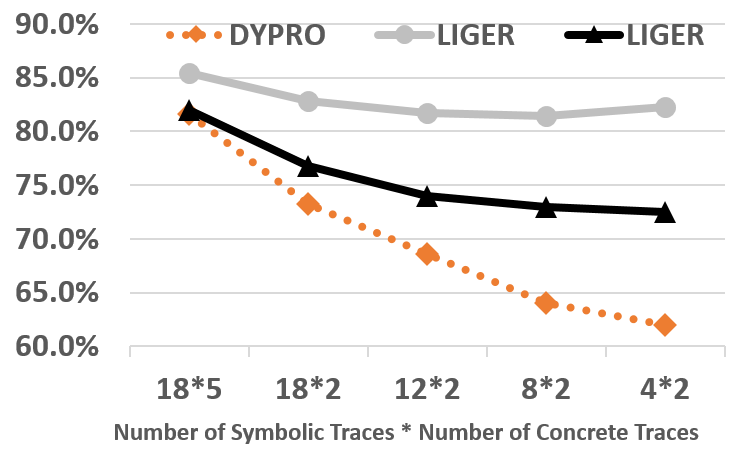}}
			\caption{Comparing \liger's accuracy trend
                          before and after removing the attention
                          mechanism in the fusion layer when both
                          symbolic and concrete traces are reduced. }
			\label{fig:abaattenchange}
		\end{center}
	\end{subfigure}
	\;
	\begin{subfigure}[b]{0.425\textwidth}
		\begin{center}
			\centerline{\includegraphics[width=\columnwidth]{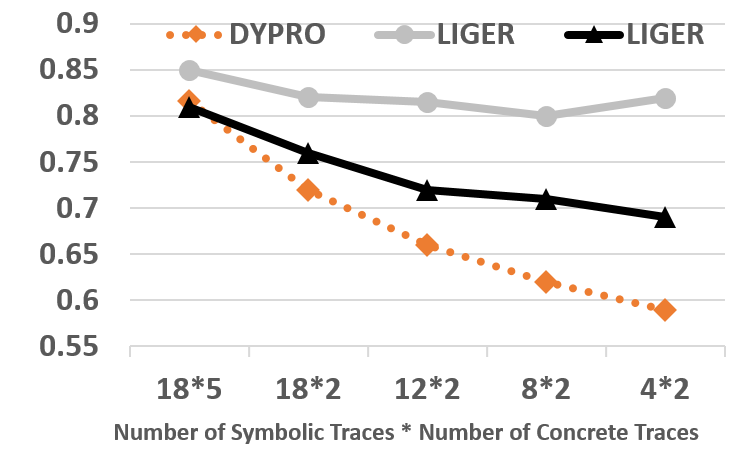}}
			\caption{Comparing \liger's F1 score trend before and after removing the attention mechanism in the fusion layer when both symbolic and concrete traces are reduced.}
			\label{fig:abaattenchangeF1}
		\end{center}
	\end{subfigure}	
	
	\caption{The effect of attention in the fusion layer on the
          model reliance on executions.
	}
	\label{fig:abaattenchangeT}
\end{figure*}

Figure~\ref{fig:abaattenT} presents the results. Removing attention
has a notable impact on \liger. Although both accuracy and F1 score
decrease, the drop is insignificant. This is an unexpected result. As
concrete traces are still abundant, an increase in their attention
weights should at least leads to a similar performance. Our hypothesis
is that allocating constant weights disrupts the balance \liger strikes for
the two feature dimensions.  Although the weights for the dynamic
feature increase, the presence of static features limits \liger's
ability to generalize.  In terms of its reliance on program
executions, \liger becomes less accurate overall
(Figure~\ref{fig:abaattenchangeT}).
The explanation is that, without the attention mechanism, 
symbolic program features will be allocated with lower weights. 
Therefore, the static feature dimension is unable to issue as 
strong signals as before to help \liger learn, thus causing 
the drop in \liger's accuracy.

\vspace{0.2cm}
\subsubsection{Summary}
\begin{figure*}[htbp!]
	\begin{subfigure}[b]{0.425\textwidth}
		\begin{center}
			\centerline{\includegraphics[width=\columnwidth]{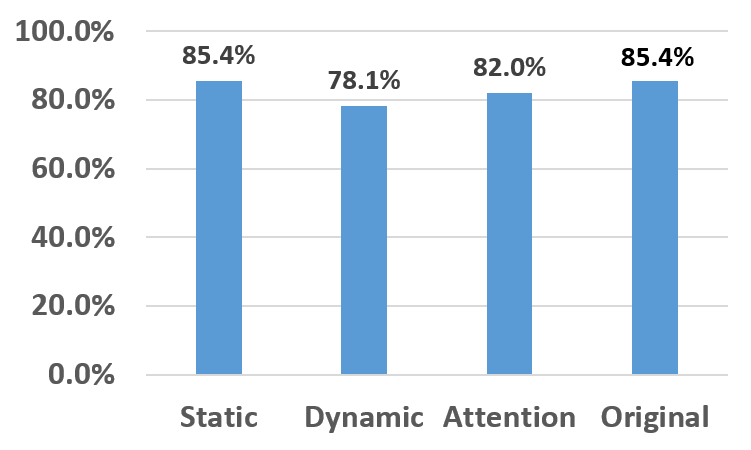}}
			\caption{Comparing \liger's accuracy in the
                          semantics classification task for each
                          ablation configuration.}
			\label{fig:comacc}
		\end{center}
	\end{subfigure}
	\;
	\begin{subfigure}[b]{0.425\textwidth}
		\begin{center}
			\centerline{\includegraphics[width=\columnwidth]{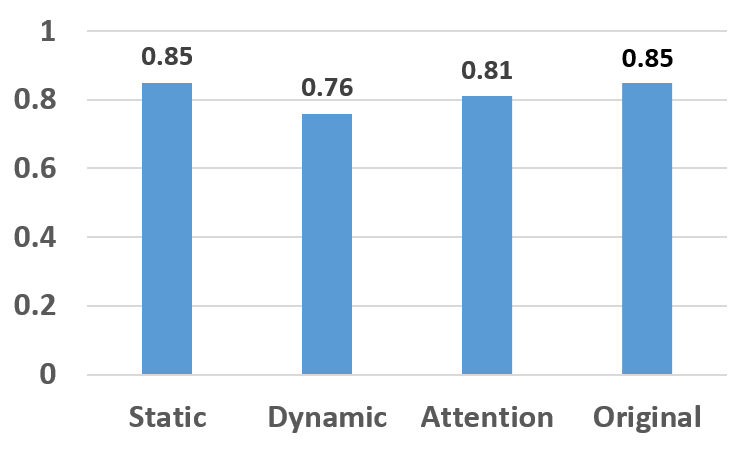}}
			\caption{Comparing \liger's F1 score in the
                          semantics classification task for each
                          ablation configuration.}
			\label{fig:comaccF1}
		\end{center}
	\end{subfigure}
	
	\begin{subfigure}[b]{0.425\textwidth}
		\begin{center}
			\centerline{\includegraphics[width=\columnwidth]{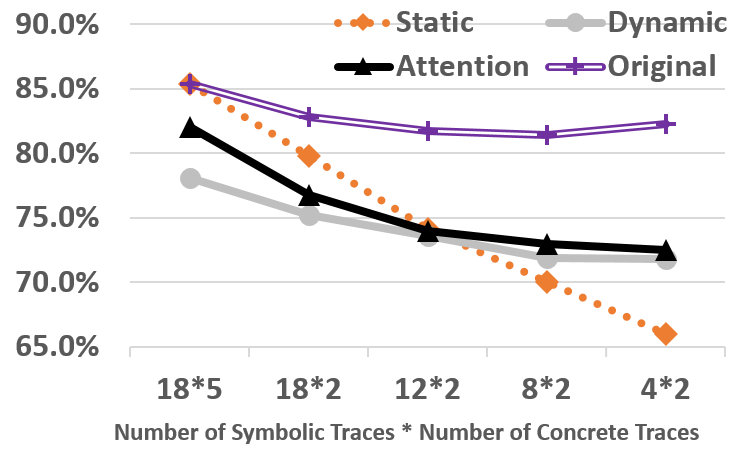}}
			\caption{Comparing \liger's accuracy trend for
                          each ablation configuration when both
                          symbolic and concrete traces are reduced but
                          branch coverage is preserved.}
			\label{fig:comre}
		\end{center}
	\end{subfigure}
	\;
	\begin{subfigure}[b]{0.425\textwidth}
		\begin{center}
			\centerline{\includegraphics[width=\columnwidth]{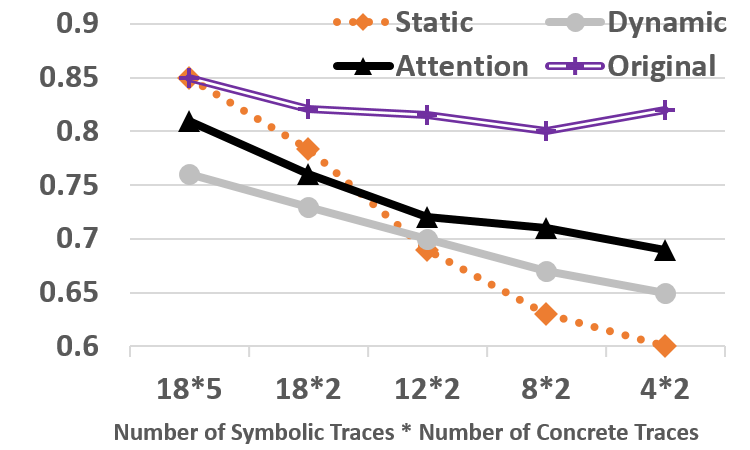}}
			\caption{Comparing \liger's F1 score trend for
                          each ablation configuration when both
                          symbolic and concrete traces are reduced but
                          branch coverage is preserved.}
			\label{fig:comreF1}
		\end{center}
	\end{subfigure}	
	
	\caption{Comparing different ablation configurations for \liger.}
	\label{fig:com}
\end{figure*}

~\\
Finally, we summarize the role of each component in the fusion layer.
To provide a direct comparison, we show the results of each new
configuration of \liger on the same diagram (Figure~\ref{fig:com}).

To summarize, \liger is the most precise model because of its
generalization from the concrete traces specially in large
quantities. However, the quality of training data is not always
guaranteed, thus programs that are difficult to cover can affect
\liger's performance. Even when code coverage is not an issue, using
an excessive amount of concrete traces leads to prolonged and
inefficient training. As a solution, combining symbolic and concrete
traces significantly reduces the reliance that \liger has on
executions in both training and testing. In addition, this feature
fusion is shown helpful in lowering \liger's need on traces with high
path coverage, provided that the original branch coverage is still
maintained for the targeted programs.
%
%
An integral part of feature fusion is the attention
mechanism which allocates greater weights to the static feature
dimension than the dynamic one. In addition, the average weight of the
static feature dimension remains largely constant when the number
of concrete traces varies. Overall, we conclude that static program
features play a more important role in feature fusion, even if they
alone are not sufficient for training a precise model to represent
program semantics.

\subsection{Method Name Prediction}

In this experiment, we examine if \liger can accurately predict the
name of a method, a problem studied
in~\cite{Alon:2019:CLD:3302515.3290353}.  Solving this task is a
strong indication of \liger's capability in learning precise
representations of program semantics.

\vspace*{3pt}
\noindent
{\textbf{Dataset.}}  We have extracted over 300K functions from a
company's database that records candidates' programs during coding
interviews. The questions asked in the coding interviews are mostly
algorithmic problems designed to evaluate a candidate's coding skills
(\eg, swapping two variables without extra memory, balancing a binary
search tree and rotating a linked list).
The function names given by the interviewers provide good descriptors
for the functions' behavior. Compared to the benchmark suite proposed
in~\cite{Alon:2019:CLD:3302515.3290353} which consists of simpler
functions (\eg, \texttt{contains}, \texttt{get} and \texttt{indexOf}),
our dataset is more challenging for a model to learn. All functions
are written in either C\# and Python. Similar to the prior
experiments, we randomly execute each program to collect the concrete
traces. By grouping concrete traces that traverse the same program
path, we derive the set of symbolic traces. We construct fifteen
blended traces, each of which is built upon five concrete traces.
Similarly, we also prepare a minimum set of blended traces that
maintain the same branch coverage as the entire set for each program
in the dataset. The minimum set contains no greater than six
traces. Each blended trace is composed of two concrete
traces. Functions that do not pass all the test cases are removed from
the dataset. In the end, we keep 174,922 functions in total which we
split into a training set of 104,922, a validation and testing set of
35K each for this experiment.

\vspace*{3pt}
\noindent
{\textbf{Evaluation Metric.}}  We have adopted the metric used by
prior work~\cite{Alon:2019:CLD:3302515.3290353} to measure precision,
recall and F1 score over case insensitive sub-tokens. The rationale is
that the prediction of a whole method name highly depends on that of the
sub-words. For example, given a method named \texttt{computeDiff}, a
prediction of \texttt{diffCompute} is considered a perfect
answer (\ie, the order of the sub-words does not matter), a prediction of
\texttt{compute} has a full precision, but low recall, and a prediction
of \texttt{computeFileDiff} has full recall, but low precision.

\vspace*{3pt}
\noindent
{\textbf{Baselines.}}  We compare \liger and code2seq on solving the
exact same tasks.  code2seq is the state-of-the-art DNN in predicting
function names. It also has a encoder-decoder architecture. The
encoder represents a function as a set of AST paths, and the decoder
uses attentions to select relevant paths while decoding.

We also include Transformer~\cite{vaswani2017attention}, arguably the
state-of-the-art deep learning model for Neural Machine Translation
(NMT).

\begin{table}[htbp!]
	\begin{center}
		\begin{adjustbox}{max width=.9\textwidth}
			\begin{tabular}{c | c | c | c } 
				\hline
				\textbf{Model} 
				& \textbf{Precision}
				& \textbf{Recall}
				& \textbf{F1 Score} \\
				\hline		
				code2seq &0.49  &0.33 &0.39 \\  									
				\hline		
				Transformer &0.46  &0.28 &0.35 \\  									
				\hline
				\hline
				\textbf{\liger}  &\textbf{0.64} &\textbf{0.52} &\textbf{0.57}\\
				\hline
				\textbf{\liger (Minimum Set)} &\textbf{0.61} &\textbf{0.44} &\textbf{0.51} \\
				\hline
			\end{tabular}
		\end{adjustbox}
	\end{center}
	\caption{Results on the models in predicting method names.}
	\label{Table:res}
\end{table}

\vspace*{3pt}
\noindent
{\textbf{Results.}}  We depict the results for each model in
Table~\ref{Table:res}. When trained on the full set of executions,
\liger significantly outperforms both competing models in all three
categories: precision, recall and F1 score. Even on the minimum set,
\liger still achieves better results than code2seq and
Transformer. Compared to Transformer, code2seq displays a slightly
better performance.  We also present the F1 score for each model as
the size of test functions increases.  \liger trained on the minimum
set of blended traces again outperforms all the other baselines across
all program lengths.

\begin{figure*}[htbp!]
	\begin{center}
		\centerline{\includegraphics[width=.65\columnwidth]{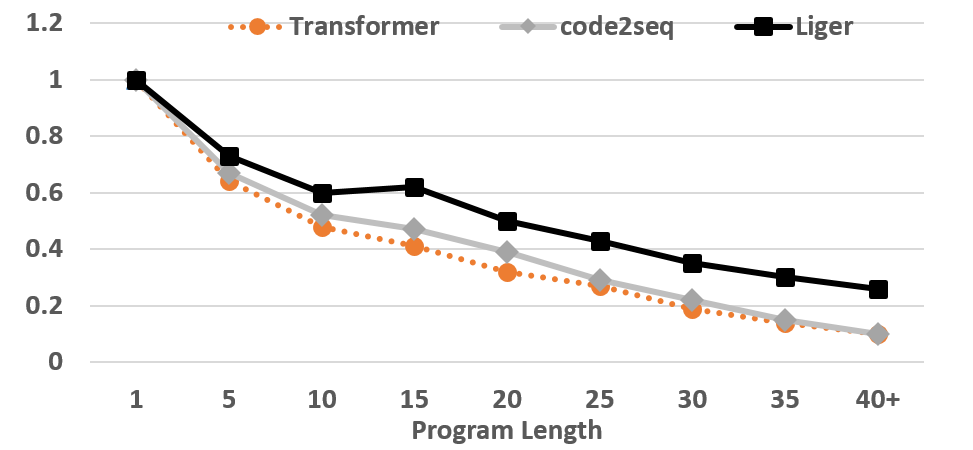}}
		\caption{Comparing models performance as the size of function increases.}
		\label{fig:namechange}
	\end{center}
\end{figure*}

\vspace*{3pt}
\noindent
{\textbf{Examples.}}  We pick two examples that \liger gives
perfect predictions, but neither code2seq nor Transformer does. The
functions are shown in Figure~\ref{fig:exanameT}.

For the function in Figure~\ref{fig:exaname1}, code2seq predicts its
name to be \texttt{swapMatrix}, while Transformer produces
\texttt{sortArray}. Although the function does exhibit the behavior of
swapping elements in a matrix, what it really does is rotating an
image 90 degrees clockwise. \liger is the only network that captures
the high-level view of the function's behavior. For the other
function, all networks successfully generate the sub-words
\texttt{FibonacciSequence}, however neither code2seq nor transformer
captures \texttt{Longest} nor \texttt{Length} indicating the
approaches' lack of precision for learning program embeddings. We have
also interacted with the code2seq tool at \url{https://code2seq.com/}
using the two functions in Figure~\ref{fig:exanameT} as inputs, and
found that the model produces incorrect results.\footnote{Interested readers
  may refer to the supplemental materials for the details.}

\begin{figure*}[htbp!]
	\vskip 0pt
	\begin{subfigure}{0.4\textwidth}
		\lstset{style=mystyle}
\begin{lstlisting}[linewidth=5.32cm]
public int rotateImage(List<List<int>> matrix)
{
    if (matrix.Count == 0)
        return -1;
    int n = matrix.Count;

    for (int i = 0; i < n; i++) {
        for (int j = 0; j < i; j++) {
            int temp = matrix[i][j];
            matrix[i][j] = matrix[j][i];
            matrix[j][i] = temp;
    } }

    for (int i = 0, j = n - 1; i < j; i++, j--) { 
        for (int k = 0; k < n; k++) { 
            int temp = matrix[k][i];
            matrix[k][i] = matrix[k][j];
            matrix[k][j] = temp;
    } }

    return 1;
}
\end{lstlisting}
		\caption{A function that rotates an image by 90 degrees (clockwise).}
		\label{fig:exaname1}
	\end{subfigure}
	\begin{subfigure}{0.5\textwidth}	
		\lstset{style=mystyle}
\begin{lstlisting}
public int LongestFibonacciSequenceLength(List<int> sequence)
{
    int length = sequence.Count;
    int longest = 0;

    for (int i = 0; i < sequence.Count - 1; i++) { 
        for (int j = i + 1; i < sequence.Count; j++) { 
            int first = sequence[i];
            int second = sequence[j];

            int count = 2;

            while (sequence.Contains(first + second)) { 
                first = second;
                second = first + second;

                count++;
                longest = Math.Max(longest, count);
    } } }

    return (longest > 2 ? longest : 0);
}
\end{lstlisting}
		\caption{A function that computes the length of the
                  longest Fibonacci sequence.}
		\label{fig:exaname2}
	\end{subfigure}
	\caption{Two example methods for which \liger correctly predicts their names.}
	\label{fig:exanameT}
\end{figure*}

%% file: related.tex
\section{Related Work}

In this section, we survey related work from three aspects: 
neural program embeddings, attention and word embeddings.

\vspace*{3pt}
\noindent
{\textbf{Neural Program Embeddings.}}  Recently, learning neural
program representations has generated significant interest in the
program languages community. The goal is to learn precise and
efficient representations to enable the application of DNNs for
solving a range of program analysis tasks. As a first step, early
methods~\cite{AAAI1714603, Pu2016,mou2016convolutional} primarily
focus on learning syntactic features. Despite these pioneering
efforts, these approaches do not precisely represent program
semantics. More recently, a number of new deep neural architectures
have been developed to tackle this
issue~\cite{allamanis2017learning,wang2017dynamic,wang2019learning,Alon:2019:CLD:3302515.3290353}. This
line of work can be divided into two categories: dynamic and
static. The former~\cite{wang2017dynamic,wang2019learning} learns from
concrete program executions, while the
latter~\cite{allamanis2017learning,Alon:2019:CLD:3302515.3290353}
attempts to dissect program semantics from source code. Unlike these
prior efforts, this paper presents an effective blended approach of
learning program embeddings from both concrete and symbolic traces.

\vspace*{3pt}
\noindent
{\textbf{Attention.}}  Attention has achieved ground-breaking results
in many NLP tasks, such as neural machine
translation~\cite{bahdanau2014neural,vaswani2017attention}, computer
vision~\cite{ba2014multiple,mnih2014recurrent}, image
captioning~\cite{xu2015show} and speech
recognition~\cite{bahdanau2016end,chorowski2015attention}.  Attention
models work by selectively choosing parts of the input to focus on while producing
the output. code2vec is among the most notable that incorporate
attention in their neural network architectures. Specifically, they
attend over multiple AST paths and assign different weights for each
before aggregating them into a program embedding. This paper uses
attention to coordinate the combination of the two feature dimensions
as well as to decode the method name as a sequence of words.

\vspace*{3pt}
\noindent
{\textbf{Word Embeddings.}}  The seminal
work~\cite{Mikolov:2013,mikolov2013efficient} of Mikolov~\etal on
word2vec stimulated the field of learning continuous
representation. They propose to embed words into a numerical space
where those of similar meanings would appear in close proximity. By
embedding words into vectors, they also discover that simple
arithmetic operations can reflect the analogies among words (\eg
$\mathit{US}-\mathit{Washington} = \mathit{China}-\mathit{Beijing}$).
word2ve, along with later efforts on learning representations of
sentences and documents~\cite{le:2014}, have greatly contributed to
state-of-the-art results in many downstream
tasks~\cite{Glorot:2011:DAL:3104482.3104547,Bengio:2003:NPL:944919.944966}.

%% file: conc.tex
\section{Conclusion}

This paper has introduced a novel, blended approach of learning
program embeddings from the combination of symbolic and concrete
execution traces. Through an extensive evaluation, we have shown that
our approach is not only the most accurate in classifying program
semantics, but also significantly outperforms code2seq, the
state-of-the-art in predicting method names. Our results have also
shown that concrete executions when supplied in large quantities
achieving good code coverage help train highly precise models.
Symbolic traces, on the other hand, reduce dynamic models' heavy
reliance on executions.  For its strong distinct benefits, we believe
that our blended approach can be adapted to tackle a wide range of
problems in program analysis and developer productivity.